\documentclass{aa}
\usepackage{graphicx}
\usepackage{natbib}
\usepackage{txfonts}
\bibpunct{(}{)}{;}{a}{}{,}
\input psfig
\begin{document}
\title{Structuring and support by Alfv\'en waves around prestellar cores}
\author{Doris Folini\inst{1} 
   \and Jean Heyvaerts\inst{1} 
   \and Rolf Walder\inst{2}}
\offprints{D. Folini}
\mail{folini@astro.u-strasbg.fr}
\institute{Observatoire de Strasbourg, 11 rue de l'Universit\'e, F-67000 Strasbourg, France;
           \and 
           Steward Observatory, University of Arizona, 933 N. Cherry Ave,
           Tucson, AZ 85721, USA; \thanks{Part of this work was done during a 6
           months research visit at the Observatoire de Strasbourg,
           France.}}
\date{Received 22 August 2002 / Accepted 1 October 2003}
\authorrunning{Folini et al.}
\titlerunning{Structuring and support by Alfv\'en waves around prestellar cores}
%
%
%
\abstract{Observations of molecular clouds show the existence of starless, dense
  cores, threaded by magnetic fields. Observed line widths indicate
  these dense condensates to be embedded in a supersonically turbulent
  environment. Under these conditions, the generation of magnetic
  waves is inevitable. In this paper, we study the structure and
  support of a 1D plane-parallel, self-gravitating slab, as a
  monochromatic, circularly polarized Alfv\'en wave is injected in
  its central plane. Dimensional analysis shows that the solution must
  depend on three dimensionless parameters. To study the nonlinear,
  turbulent evolution of such a slab, we use 1D high resolution
  numerical simulations. For a parameter range inspired by molecular
  cloud observations, we find the following.  1) A single source of
  energy injection is sufficient to force persistent supersonic
  turbulence over several hydrostatic scale heights.  2) The time
  averaged spatial extension of the slab is comparable to the
  extension of the stationary, analytical WKB solution.  Deviations,
  as well as the density substructure of the slab, depend on the
  wave-length of the injected wave. 3) Energy losses are dominated by
  loss of Poynting-flux and increase with increasing plasma beta. 4)
  Good spatial resolution is mandatory, making similar simulations in
  3D currently prohibitively expensive.   }
\maketitle
\section{Introduction}
\label{sec:intro}
Magnetic fields are observed in at least some molecular clouds
\citep{1999ApJ...520..706C,bourke-et-al:01}. Whether all molecular
clouds are threaded by magnetic fields is still under debate. 
\citet{2000ApJ...537L.135W} observe ordered magnetic fields on small scales of about 
0.05 pc in, adopting their terminology, prestellar cores ($N \approx
10^{5}$ cm$^{-3}$). Also on somewhat larger scales, in star forming
regions, ordered magnetic fields are
reported~\citep{2002ApJ...571..356M,2002ApJ...569..304M}.  Coherent
velocities in prestellar cores are observed on scales of about 0.01
pc~\citep{1998ApJ...504..207B}. On larger scales, observed line widths
indicate supersonic motions. Taken together, these observations
suggest dense condensates, threaded by magnetic fields, to be
embedded in a supersonically turbulent environment. The generation of
magnetic waves under such conditions is inevitable.

On larger scales, such magnetic waves are likely to be strongly damped
(e.g. by ion-neutral friction or instabilities) or dominated by other
processes (e.g. ISM-turbulence or incoming magnetic waves as studied
by~\citet{elmegreen:99}). This finding is in agreement with molecular
cloud theories and observations. While some years ago it was thought
that molecular clouds had to be supported against their self-gravity
for at least $10^{8}$ years, new results are much more in agreement
with a picture in which molecular clouds form, stars are born, and the
clouds are dispersed, all within some $10^{6}$ years. Observations of
molecular clouds in the solar neighborhood show that most clouds do
form stars~\citep{2001ApJ...562..852H}, from which it is concluded
that star formation begins essentially as soon as a molecular cloud
forms.  Using stellar evolutionary tracks leads to the further
conclusion that star formation in a molecular cloud takes place
rapidly, once it has started~\citep{palla-stahler:00}. For stellar
populations with an average age larger than about 3 Myr, no more
molecular material can be detected \citep[][and references
therein]{2001ApJ...562..852H}, indicating that star formation also
ceases rapidly.  Numerical simulations also support such a dynamical
scenario~\citep{1999ApJ...527..285B,2000ApJ...530..277E,maclow:02}.

For smaller spatial scales, on the other hand, recent observations and
simulations support the idea that magnetic fields and waves play an
important role in the structuring of the environment of -- possibly
only transient -- high density molecular clumps and the inhibition of
accretion onto such clumps.  Observations of the starless dense core
L1512 and its immediate vicinity by~\citet{2001ApJ...555..178F} show
six dense filaments pointing towards the core and extending up to
about 1 pc.  The matter within each filament is observed to move
towards the core while probably describing circular motions in the
direction transverse to the filament. This motion and the orientation
of the filaments make it likely that they are not merely part of the
turbulent cascade within the cloud.

\citet{2001ApJ...547..280H} performed grid studies for 3D hydrodynamical and
MHD simulations of the formation of persistent cores. They find that
increasing spatial resolution leads to increased accretion in the
hydrodynamical case, but to a decreased one in the MHD case. The
authors ascribe this difference to better resolution of MHD waves,
which then counteract accretion. For the case of a plane-parallel
slab, at the boundary of which a monochromatic Alfv\'en wave is
injected, 2D simulations with constant gravity
by~\citet{pruneti-velli:97}, as well as 2D and 3D simulations without
gravity by~\citet{del-zanna-et-al:01}, show the development of
high-density filaments parallel to the direction of propagation of the
Alfv\'en wave.  Note that filamentation is observed only when open,
not periodic, boundaries are used at the planes perpendicular to the
direction of wave-propagation. \citet{2002ApJ...566L..49C} have shown
that magnetic fields can have rich structures well below the viscous
dissipation scale, possibly affecting the density structure as well.

In this paper, we study the effect of magnetic waves in the framework
of a simplified model, a 1D plane-parallel, self-gravitating slab with
one central source of monochromatic, circularly polarized Alfv\'en
waves.  Related models have been investigated by other authors
before~\citep{gammie-ostriker:96,
martin-et-al:97,1999ApJ...517..226F,2002MNRAS.329..195F,
kudoh}\footnote{The paper by Kudoh \& Basu was submitted during the
refereeing process of this paper.}. The work here differs from
previous 1D simulations in that we focus on the highly nonlinear,
long-term evolution. Also, we consider only one central source of
Alfv\'en waves, instead of injecting energy at each grid point, and we
use open boundaries, not periodic ones.  For this setting, we present
a dimensional analysis as well as a parameter study based on numerical
simulations.

Our results show that already one source of waves is sufficient to
structure and support a turbulent slab in a quasi-static manner.  The
WKB solution for an Alfv\'{e}n-wave supported, self-gravitating 1D
slab by~\citet{martin-et-al:97} gives a good order of magnitude
estimate for the average spatial extent of the turbulent slab, but
fails to account for the rich interior structure.  And while the
Poynting-flux is constant in the WKB solution, there is substantial
loss of Poynting-flux in the solution of the full equations. As
governing parameters for deviations of the numerical solution from the
analytical WKB solution we identify the initial, central Alfv\'en
wave-length and the initial, central plasma beta.  This is remarkable
in view of the highly nonlinear, turbulent nature of the slab, where,
for example, the true central Alfv\'en wave-length loses, even on
average, any connection with its initial value after a fraction of a
free-fall time. To observe both the structuring and support of the
slab by magnetic waves, we find a good spatial resolution and high
order of integration to be decisive.

The paper is organized as follows. In Sect.~\ref{sec:model} we
describe our physical model and the numerical method we use.  We give
a dimensional analysis of the problem in Sect.~\ref{sec:dimensionless}
before proceeding to the numerical results in
Sect.~\ref{sec:num_results}.  A discussion of our results follows in
Sect.~\ref{sec:discussion}, conclusions are given in
Sect.~\ref{sec:conclusions}.
\section{The model}
\label{sec:model}
\subsection{Physical model problem}
\label{sec:phys_mod}
We consider a 1D (x-direction), plane-parallel, self-gravitating slab
which we assume to be symmetric with respect to a central plane at
$x=0$ (yz-plane, infinitely extended), where an Alfv\'{e}n-wave is
injected. In this geometry, all variables are functions of distance
$x$ to the central plane and time $t$ only. Velocities and magnetic
fields perpendicular to the x-direction are allowed, but gradients can
occur only in the x-direction. To describe the time evolution of this
slab, we use the ideal, isothermal MHD equations, including a source term
to account for self-gravity. In their conservative formulation, these
equations read:
\begin{eqnarray}
\label{eq:div}
\frac{\partial \rho}{\partial t} + 
      \nabla \cdot (\rho \vec{v}) & = & 0, \\
\label{eq:mom}
\frac{\partial (\rho \vec{v})}{\partial t} + 
     \nabla \cdot \left[\rho \vec{v} \vec{v} + 
                        p_\mathrm{tot}\vec{I} - 
                        \frac{\vec{B}\vec{B}}{\mu_\mathrm{0}}
                  \right] & = & \rho  \vec{g}, \\
\label{eq:mag}
\frac{\partial \vec{B}}{\partial t} +
     \nabla \cdot (\vec{v}\vec{B} - \vec{B}\vec{v}) & = & 0. 
\end{eqnarray}
Here, $\rho$ denotes the mass density, $\vec{v} =
(v_\mathrm{x},v_\mathrm{y},v_\mathrm{z})$ the velocity, and
$\vec{B} = (B_\mathrm{x},B_\mathrm{y},B_\mathrm{z})$ the magnetic
field. We will often use the notation $B_\mathrm{\parallel}$ for the
constant background magnetic field $B_\mathrm{x}$, as well as
$B_\mathrm{\perp}$ for the magnitude of the transverse magnetic field,
and $B_\mathrm{\perp 0}$ for $B_\mathrm{\perp}$ at $x=0$ and $t=0$.
Generally, a subscript $0$ to a variable refers to the initial
($t=0$) and, if space dependent, central ($x=0$) value of a
variable.  $\mu_\mathrm{0}$ is the magnetic permeability of free
space, and $\vec{I}$ is the identity tensor.  $p_\mathrm{tot} =
p_\mathrm{th} + (1/2\mu_\mathrm{0})\vec{B}^{2}$ denotes the total
pressure, where the thermal pressure is given by the isothermal
equation of state $p_\mathrm{th} = \rho c_{\mathrm{s}}^{2}$, with
$c_{\mathrm{s}} =
\sqrt{RT}$ the isothermal sound speed. $T$ is the temperature of the
gas and $R$ is the gas constant. Within the framework of our 1D model,
the force exerted by self-gravity in the x-direction is given by
$\vec{g} = (g(x,t),0,0)$ with
\begin{equation}
g(x,t) = - 4 \pi G \int_{0}^{x} \rho(x',t) dx'.
\label{eq:sg}
\end{equation}
$G$ denotes the gravitational constant.  We consider only one sort of
particle and consequently have no wave damping due to ion-neutral
friction in our model. The average mass per particle is the mass
$m_\mathrm{H}$ of a hydrogen atom.

For special cases, analytical solutions exist. For an infinitely
extended slab with $\vec{B}=\vec{0}$, the stationary solution of the
above model problem is given by the hydrostatic density distribution
\citep{spitzer:68},
\begin{equation} 
\rho(x) = \rho_\mathrm{0} \cdot \frac{1}{\cosh^{2}(x/H)}, 
\label{eq:hydstat} 
\end{equation} 
where $\rho_\mathrm{0}$ is the mass density at the central plane of
the slab, $H = c_{\mathrm{s}}^{2} / (2 \pi G {\cal
M}^{\mathrm{hs}})$ is the hydrostatic scale height, and
${\cal M}^{\mathrm{hs}} = \int_{0}^{\infty}
\rho(x) dx$ is the column density of the hydrostatic slab.  
$\rho_{\mathrm{0}}$ and $c_{\mathrm{s}}$ are free parameters of the
solution.

For the case where the magnetic background field $B_\mathrm{\parallel}
\ne 0$, and when a monochromatic, circularly polarized  Alfv\'en wave 
is injected at the central plan of the slab, \citet{martin-et-al:97}
derived a stationary, analytical solution of the above model problem
in the framework of a WKB approximation. In contrast to the hydrostatic
solution, the WKB solution has three free parameters:
$\rho_{\mathrm{0}}$, $c_{\mathrm{s}}$, and $B_{\mathrm{\perp}}$.
\subsection{Model parameters and naming conventions}
\label{sec:params}
The full model problem as formulated in Sect.~\ref{sec:phys_mod} has
five free parameters. They could be specified in a dimensionless form,
as we are going to discuss in Sect.~\ref{sec:dimensionless}.  Guided
by observations of molecular clouds, we choose, however, the following
set of dimensional parameters: the background magnetic field
$B_{\mathrm{\parallel}}$, the temperature $T$ of the slab, the
initial, central mass density $\rho_{\mathrm{0}}$, the amplitude of
the Alfv\'{e}n wave, specified by either $B_{\mathrm{\perp} 0}$ or
$v_{\mathrm{\perp} 0}=B_{\mathrm{\perp} 0}/\sqrt{\mu_{\mathrm{0}}
\rho_{\mathrm{0}}}$, and the frequency $\omega$ of the wave.

We varied parameters within limits that correspond roughly to observed
parameters in molecular clouds: magnetic fields of 10-100 $\mu$G,
temperatures between 5 K and 40 K, and central particle densities ranging from
250 to 2000cm$^{-3}$. For the wave frequency $\omega$ we
have assumed values in the range $ 10^{4} \mbox{yr} \le 2\pi/\omega
\le 2.5 \cdot 10^{5}
\mbox{yr}$. With this choice of parameters we are in a low-beta
regime, the initial, central plasma beta $\beta_{0} = 2
c_{\mathrm{s}}^{2} / c_\mathrm{A0}^{2}$ lying in a range between 0.003
and 0.7.  Here, $c_{\mathrm{A0}} =
B_{\mathrm{\parallel}}/\sqrt{\mu_{\mathrm{0}}\rho_{\mathrm{0}}}$
denotes the initial, central Alfv\'{e}n-speed. The corresponding 
Alfv\'{e}n wave-length $\lambda_{\mathrm{A0}} = c_\mathrm{A0}
\cdot 2\pi/\omega$, lies in a range between about 0.07 pc and 0.35 pc.
The magnetic field $B_{\mathrm{\perp} 0}$ corresponds to
transverse velocities $v_{\mathrm{\perp} 0}$ in the range between $2.5
\cdot 10^{4}$cm/s and $3.2 \cdot 10^{5} $cm/s.
The detailed parameters for each of the performed simulations are
given in Table~\ref{tab:mod_params} of the appendix. They are also
reflected in the name of each simulation. For example, R20.10.25.4.40
is the simulation with $B_{\mathrm{\parallel}}=20\mu$G,
$B_{\mathrm{\perp}}=10\mu$G, $2\pi/\omega= 25 \times 10^{4}$ years, a
central density of $N_{\mathrm{0}} = 4 \times 250$ particles per
cm$^{3}$, and $T=40$K.
\subsection{Numerical solution}
\label{sec:numerical_method}
\subsubsection{Numerical method}
In our simulations, we consider a slab of finite (not
infinite) spatial extension ${\cal D}$.  We use a finite volume method
on an equidistant spatial grid to solve the ideal, isothermal MHD
equations, Eqs.~\ref{eq:div}--\ref{eq:mag}. Fluxes are computed using
a second order in time and third order in space stabilized
Lax-Friedrichs solver\footnote{The code is part of the A-MAZE code
  package \citep{amaze:00}, comprising 3D adaptive mesh codes for
  magneto-hydrodynamics and radiative transfer. The codes are
  available at http://www.astro.phys.ethz.ch/staff/folini.} as
described in \citet{barmin-et-al:96}. As is shown in
\citet{barmin-et-al:96}, the accuracy of this solver is comparable to
that of a second order Riemann solver. Self-gravity is taken into
account using a Strang-splitting \citep{strang:68}.
\subsubsection{Initial conditions}
\label{sec:initial_conditions}
At time $t=0$ we assume the slab to have a density distribution
according to the analytical WKB solution of \citet{martin-et-al:97}
for a given set of parameters as specified in
Table~\ref{tab:mod_params}. The velocity in the x-direction, as well as
the transverse components of the velocity and the magnetic field, we
set to zero.  The magnetic field in x-direction,
$B_{\mathrm{\parallel}}$, and the temperature $T$, we set to the
values given in Table~\ref{tab:mod_params}. The mass column density of
the initial WKB solution, within the computational domain and to
infinity, we denote by ${\cal M}_{\mathrm{0}}$ and ${\cal
M}_{\mathrm{0}}^{\mathrm{\infty}}$ respectively.
\subsubsection{Boundary conditions}
\label{sec:boundary_conditions}
Boundary conditions are implemented using four boundary cells at each
of the two domain boundaries. Note that in this way we merely control
the physical variables set in these cells. The fluxes entering
and leaving the domain are determined by the interaction of the solution
as set in the boundary cells and the numerical solution within the
computational domain.

In the boundary cells at the inner domain boundary $(x=0)$, the physical
variables are set in accordance with a left-handed, circularly
polarized Alfv\'{e}n-wave, whose velocity amplitude $v_\mathrm{\perp}$
we keep fixed in time, $v_{\mathrm{\perp}} = v_{\mathrm{\perp
0}}$. Allowing $v_{\mathrm{x}}$ and $\rho_{\mathrm{0}}$, and thus
$B_{\mathrm{\perp}}$, to vary according to the solution in the domain,
this yields:
\begin{eqnarray}
\rho(x)         & = &   \rho(-x),  \nonumber \\ 
v_\mathrm{x}(x) & = & - v_\mathrm{x}(-x), \nonumber \\ 
v_\mathrm{y}(x) & = &   v_\mathrm{\perp} \cos(\omega(t - \Delta t)), \nonumber  \\
v_\mathrm{z}(x) & = &   v_\mathrm{\perp} \sin(\omega(t - \Delta t)), \nonumber \\ 
B_\mathrm{x}(x) & = &   B_\mathrm{\parallel},  \nonumber \\ 
B_\mathrm{y}(x) & = &   v_\mathrm{y}(x) \cdot \sqrt{\mu_\mathrm{0}\rho(x)}, \nonumber \\ 
B_\mathrm{z}(x) & = &   v_\mathrm{z}(x) \cdot \sqrt{\mu_\mathrm{0}\rho(x)}.  
\end{eqnarray}
For the calculation of the time retardation $\Delta t =
x/c_{\mathrm{A}}$ we take into account that $c_\mathrm{A}$ varies with
$x$ and $t$. 

At the outer boundary, we distinguish two cases.  If $v_{x}({\cal
D},t)>0$ we use a zeroth order extrapolation for all conserved
variables.  If $v_{x}({\cal D},t)<0$ we use a zeroth order extrapolation
for $\vec{v}$ and $\vec{B}$, but restrict the density to $10^{-4}
\rho_\mathrm{0}$. Note that these outer boundary conditions allow for
both accretion or loss of matter, energy, and momentum. Associated
changes in the mass column density over the domain, ${\cal M}(t)$,
are, however, mostly less than 2\% (see Table~\ref{tab:mod_params}).

We have chosen open boundaries as these match best with our intention
to investigate the effect of only one source of Alfv\'en waves. Using
periodic boundary conditions instead would implicitly introduce
several sources of Alfv\'en waves, separated from each other by a
distance $2{\cal D}$.

\subsubsection{Choice of domain size, discretization, and integration time}
\label{sec:choice}
For the simulations we chose a domain of size ${\cal D}=6$ pc, or
about 20 hydrostatical scale heights, covered by 5000 cells. With this
choice, we fulfill the following four basic requirements. 1) The
initial WKB solution fits well on the domain, 90\% of its column
density ${\cal M}_{\mathrm{0}}$ occupy less than 60\% of the domain.
2) The numerical solution fits well on the domain, 90\% of its
column density ${\cal M}(t)$ are contained in the inner half of the
domain and ${\cal M}(t)$ barely changes with time. 3) The hydrostatic
solution is covered by sufficiently many cells. 4) The numerical
solution does not depend on the discretization. Several grid studies
show 5000 cells to be both necessary and sufficient (see
Sect.~\ref{sec:disc_num} and Fig.~\ref{fig:spaceres}).  We followed
all our simulations for $2
\cdot 10^{7}$ years, or about 20 sound crossing times of the
hydrostatic scale height.
\section{Dimensional analysis}
\label{sec:dimensionless}
Before coming to the numerical results in Sect.~\ref{sec:num_results},
we present a dimensional analysis of the model problem formulated in
Sect.~\ref{sec:phys_mod}.  Dimensional parameters which completely
specify the problem are the initial mass column density ${\cal
M}_{\mathrm{0}}^{\mathrm{\infty}}$, the magnetic field
$B_{\mathrm{\parallel}}$, the sound speed $c_{\mathrm{s}}$, the wave
frequency $\omega$ of the imposed oscillations, and their velocity
amplitude $v_{\mathrm{\perp} 0}$ at the central plane. Since Alfv\'en
wave propagation and self-gravity are essential parts of the problem,
the dimensional constants $\mu_{\mathrm{0}}$ and $G$ also determine
the solution.  The solution in infinite space is determined by these
seven dimensional input parameters. For our finite computational
domain this is no longer strictly true, but we shall neglect this
complication for the moment.

From the seven dimensional input parameters we can build four natural
reference values for length, time, mass density, and magnetic field.
The three extra dimensional parameters would then be associated with
three dimensionless quantities that can be constructed from the seven
parameters.  In infinite space, the (suitably normalized) physical
quantities in the solution would be functions of the normalized time
and space variables and of these three dimensionless input parameters.
\subsection{Natural dimensional reference values}
\label{sec:naturaldimles}
The four natural dimensional reference values (subscript $\mathrm{u}$
for 'unity' in the following) should be defined such that none of them
would approach zero or infinity when some parameters of the problem
take expectedly large or small values. For example, we should allow
$\omega$ to become very large, as it might be in the WKB limit and as
it is possibly met in actual situations. Similarly, $c_{\mathrm{s}}$
may, in some clouds, become small enough to be neglected.  This means
that sensible natural references should be constructed from ${\cal
M}_{\mathrm{0}}^{\mathrm{\infty}}$, $B_{\mathrm{\parallel}}$,
$v_{\mathrm{\perp} 0}$, $\mu_{\mathrm{0}}$ and $G$ alone.  Obviously
$B_{\mathrm{\parallel}}$ provides a reference magnetic field,
$B_{\mathrm{u}}$, while the references of mass density, length, and
time, $\rho_{\mathrm{u}}$, $d_{\mathrm{u}}$ and $t_{\mathrm{u}}$, can
be defined from ${\cal M}_{\mathrm{0}}$, $v_{\mathrm{\perp} 0}$ and
$G$ alone.  Straightforward dimensional analysis shows that these four
reference scales can be taken as:
\begin{eqnarray}
B_{\mathrm{u}}    & = & B_{\mathrm{\parallel}}, \nonumber \\
\rho_{\mathrm{u}} & = & \frac{ 2 \pi G ({\cal M}_{\mathrm{0}}^{\mathrm{\infty}})^2}
                             {v_{\mathrm{\perp} 0}^2}, \nonumber \\
d_{\mathrm{u}}    & = & \frac{ v_{\mathrm{\perp} 0}^2}
                             {2 \pi G {\cal M}_{\mathrm{0}}^{\mathrm{\infty}}}, \nonumber \\
t_{\mathrm{u}}    & = & \frac{ v_{\mathrm{\perp} 0}}
                             {2 \pi G {\cal M}_{\mathrm{0}}^{\mathrm{\infty}}}.  
\label{eq:nondim1}
\end{eqnarray}
The factors $2\pi$ were inserted for convenience, such that
$\rho_{\mathrm{u}}$ and $d_{\mathrm{u}}$ are the central density and
scale height of a self-gravitating isothermal sheet with sound
speed $v_{\mathrm{\perp}}$ that would have, in infinite space, a 
column density ${\cal M}_{\mathrm{0}}^{\mathrm{\infty}}$.  The
reference time $t_{\mathrm{u}} = d_{\mathrm{u}}/v_{\mathrm{\perp} 0}$ is
of the order of the Jeans
period associated with $\rho_{\mathrm{u}}$.  

From the three remaining dimensional input parameters $\mu_{\mathrm{0}}$,
$c_{\mathrm{s}}$ and $\omega$ we can form, given these references, 3
natural dimensionless numbers, for example:
\begin{eqnarray}
\alpha_{\mathrm{u}} & = & 
      \frac{ v_{\mathrm{\perp} 0}^2}
           {\left( B_{\mathrm{u}}^2/(\mu_{\mathrm{0}} \rho_{\mathrm{u}})\right)}, \nonumber \\
\beta_{\mathrm{u}}  & = &  
      \frac{ 2 c_S^2}
           { \left( B_{\mathrm{u}}^2/(\mu_{\mathrm{0}} \rho_{\mathrm{u}})\right)}, \nonumber \\
W_{\mathrm{u}} & = & 
      \frac{\omega }{ \sqrt{ 2 \pi G \rho_{\mathrm{u}}} }.
\label{eq:nondim2}
\end{eqnarray}
The parameter $\alpha_{\mathrm{u}}$ compares the imposed velocity
amplitude $v_{\mathrm{\perp}0}$ to the reference Alfv\'en velocity
$c_{\mathrm{Au}}$, defined by $c_{\mathrm{Au}}^2 = B_{\mathrm{u}}^2
/(\mu_{\mathrm{0}} \rho_{\mathrm{u}})$.  The parameter $
\beta_{\mathrm{u}}$ is the ratio of the reference gas pressure
to the magnetic pressure of $B_{\mathrm{u}}$. The parameter
$W_{\mathrm{u}}$ is a WKB parameter. Clearly, neither the set of the
four reference dimensional values nor the set of the three input
dimensionless numbers is uniquely defined.  Other dimensional
reference values could be obtained by multiplying the chosen ones by
any function of the dimensionless input numbers $\alpha_{\mathrm{u}}$,
$\beta_{\mathrm{u}}$, and $W_{\mathrm{u}}$.  Similarly, the above set
of three such numbers could be replaced by three other arbitrary
functions of them.
\subsection{Dimensional reference values from WKB solution}
\label{sec:wkbdimles}
The reference quantities defined in Eq.~\ref{eq:nondim1} need not be
very close to the actual values of the physical quantities in the
solution, even near the central plane.  It may be felt desirable,
however, that the reference values be as close as possible to actual
ones, even at the price of defining the reference values in a more
sophisticated way than $\rho_{\mathrm{u}}$, $d_{\mathrm{u}}$ and
$t_{\mathrm{u}}$. If the WKB solution is to be a guide, such
estimates could be obtained from the solution of Martin et
al. (1997). From this solution, it is possible to relate the central
density $\rho_{\mathrm{0}}$ of a solution to its column density ${\cal
M}_{\mathrm{0}}$ in the finite computational domain of thickness
${\cal D}$ and to the total column density to infinity, ${\cal
M}_{0}^{\mathrm{\infty}}$.  The limited extent of the computational
domain is reflected in the fact that the ratio $({\cal
M}_{\mathrm{0}}/{\cal M}_{\mathrm{0}}^{\mathrm{\infty}})$ is less than
unity.  A scale length $d^{\mathrm{wkb}}_{\mathrm{0}}$ may then be
defined as the distance from the central plane of the slab where 90\%
of the column density ${\cal M}_{\mathrm{0}}$ are reached. The time scale
$t_{\mathrm{0}}$ can be defined as $t_{\mathrm{0}}= (2 \pi G
\rho_{\mathrm{0}})^{-1/2}$. The reference value of the magnetic field may be
chosen as before, $B_{\mathrm{0}} = B_{\mathrm{\parallel}}$.

In analogy with Eq.~\ref{eq:nondim2}, dimensionless numbers
$\alpha_{\mathrm{0}}$, $\beta_{\mathrm{0}}$, and $W_{\mathrm{0}}$ can
be introduced. Note that $\beta_{\mathrm{0}}$ is, in fact, identical
to $\beta_{\mathrm{0}}$ as given in Sect.~\ref{sec:params}. An
Alfv\'{e}n-velocity can be defined by $c_{\mathrm{A0}}^{2} =
B_{\mathrm{\parallel}}^2 / (\mu_{\mathrm{0}}\rho_{\mathrm{0}})$, and
Alfv\'{e}n wave-length by $\lambda_{\mathrm{A0}} = 2 \pi
c_{\mathrm{A0}}/\omega$.

Aliases to the WKB parameter $W_{\mathrm{0}}$ could be used as
well. In particular, one could prefer the ratio of the WKB length
scale to the Alfv\'en wavelength associated with
$B{\mathrm{\parallel}}$ and $\rho_{\mathrm{0}}$, $W_{\mathrm{\lambda
A0}} = \lambda_{\mathrm{A0}}/d^{\mathrm{wkb}}_{\mathrm{0}}$.  For our
simulations, the values for a number of these dimensionless parameters
are listed in Table~\ref{tab:mod_params}.
\subsection{Reference values for substructure scale}
\label{sec:refsub}
Unlike the WKB solution, the numerical solution of the model problem
from Sect.~\ref{sec:phys_mod} is far from smooth (see
e.g. Fig.~\ref{fig:r20.20.25.4}). And while the lengths
$d_{\mathrm{u}}$ or $d^{\mathrm{wkb}}_{\mathrm{0}}$ are natural
estimates for the global size of the mass distribution, the natural
scale length associated with substructures induced by the wave is, of
course, quite different. It is either of the order of the reference
Alfv\'en wavelength $\lambda_{\mathrm{Au}} = 2 \pi
c_{\mathrm{Au}}/\omega$ or of the Alfv\'en wavelength associated with
the WKB reference quantities, $\lambda_{\mathrm{A0}} = 2 \pi
c_{\mathrm{A0}}/\omega$. These two scales are of the same order of
magnitude for a wave-supported cloud.  The actual characteristic scale
of cloud substructure is the product of any one of them with a
function of the dimensionless input parameters.  On physical grounds,
we expect, however, this function to be of the order of unity, since no other
small length scale is likely to play an important role.  In the
low-$\beta$ regime we consider, the sonic wavelength $2 \pi c_S
/\omega$ is usually much shorter and sound perturbations nonlinearly
evolve into shocks anyway, so that this sonic wavelength is not
expected to show up in the spectrum of the fluctuations, except
perhaps at the level of the very small thickness of the dense sheets
that form.

In the next section, we present the numerical solutions for the
model problem from Sect.~\ref{sec:phys_mod} for various parameter
sets.  Note that our study, inspired by molecular clouds, covers only
a small part of the entire parameter space. For this part of the
parameter space, we identify and discuss some of the dependences
indicated by the dimensional analysis.
\section{Numerical results}
\label{sec:num_results}
The injection of magnetic waves at the central plane of the
self-gravitating slab has two major consequences: the
slab becomes supersonically turbulent and its spatial extension is
clearly larger than in hydrostatic equilibrium. 

\begin{figure}[htp]
\centerline{
\includegraphics[width=8.5cm]{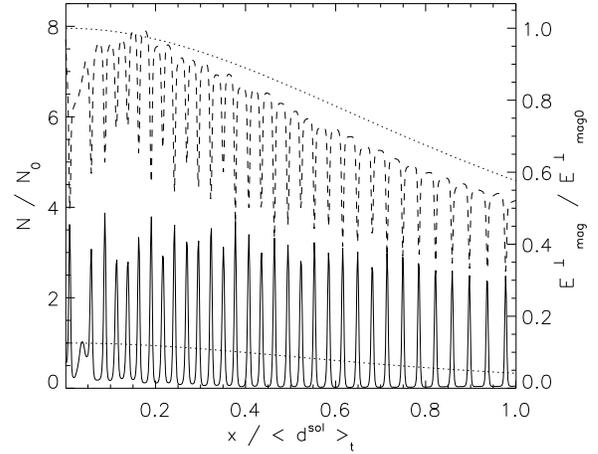}
           }
\caption{Early on in all our simulations, high density sheets develop
         under the influence of the parametric instability of the
         Alfv\'{e}n-wave, while the transverse magnetic field is
         partly destroyed. The figure is a snapshot of simulation
         R20.20.5.4.10 after $2.1 \cdot 10^{5}$ years. Shown are the
         density (solid line) and the energy density of the transverse
         magnetic field (dashed line), normalized to their initial,
         central values, as functions of distance to the central plane
         (x-axis, in units of $\langle d^{\mathrm{sol}}
         \rangle_{\mathrm{t}}$, the time averaged slab extension). The
         dotted curves denote $N/N_{\mathrm{0}}$ (lower curve) and
         $E^{\mathrm{\perp}}_{\mathrm{mag}} /
         E^{\mathrm{\perp}}_{\mathrm{mag0}}$ (upper curve) of the
         initial WKB solution.}
\label{fig:param}
\end{figure}

Neither of these consequences is surprising. The energy provided by
the injected wave must result in additional support of the slab
against its self-gravity. Density inhomogeneities are to be
expected since within the frame of linear analysis a parametric
instability of the injected Alfv\'en wave exists
\citep{1978ApJ...224.1013D, 1978ApJ...219..700G, turkmani}. Early on in our
simulations we observe the associated growth of high density sheets,
which is accompanied by the destruction of the transverse magnetic
field (see Fig.~\ref{fig:param}). Note also that
\citet{malara-velli:96} and \citet{malara-et-al:00} demonstrated that
even a non-monochromatic spectrum of Alfv\'en waves is subject to
parametric instability with linear and nonlinear growth rates of the
same order of magnitude as in the monochromatic case.

So far the turbulent, nonlinear evolution of such a system is,
however, not well investigated.  The results we present in the
following demonstrate that the same parameters which govern the
parametric instability, $\lambda_\mathrm{A0}$ and $\beta_\mathrm{0}$,
are also of crucial importance for the nonlinear, turbulent solution.
This despite the fact that the true, time averaged values of these
quantities at $x=0$ (and close by) deviate substantially from
$\lambda_\mathrm{A0}$ and $\beta_\mathrm{0}$.
\begin{figure*}[htp]
\centerline{
\includegraphics[width=8.5cm,height=6.9cm]{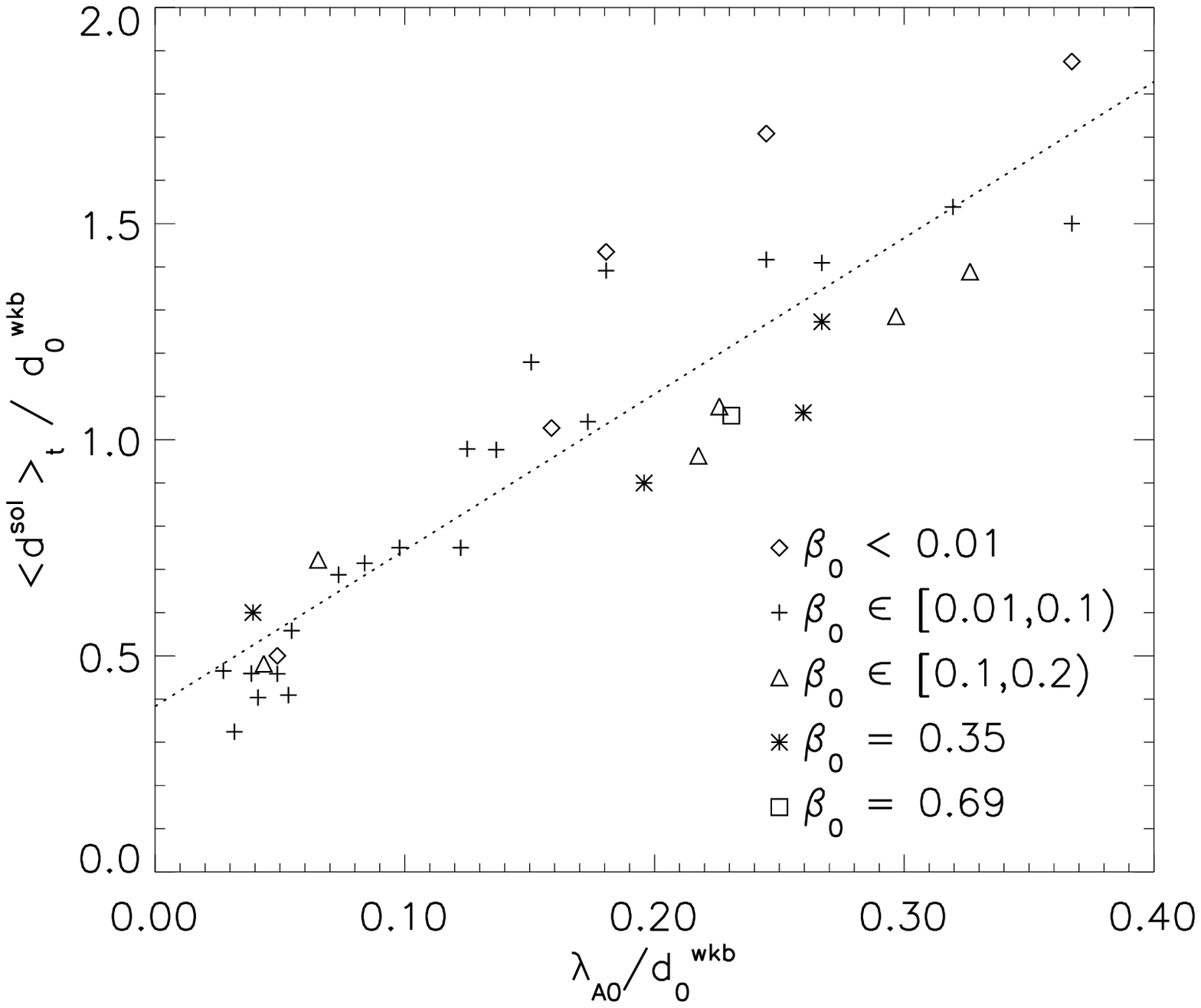}
\includegraphics[width=8.1cm,height=6.9cm]{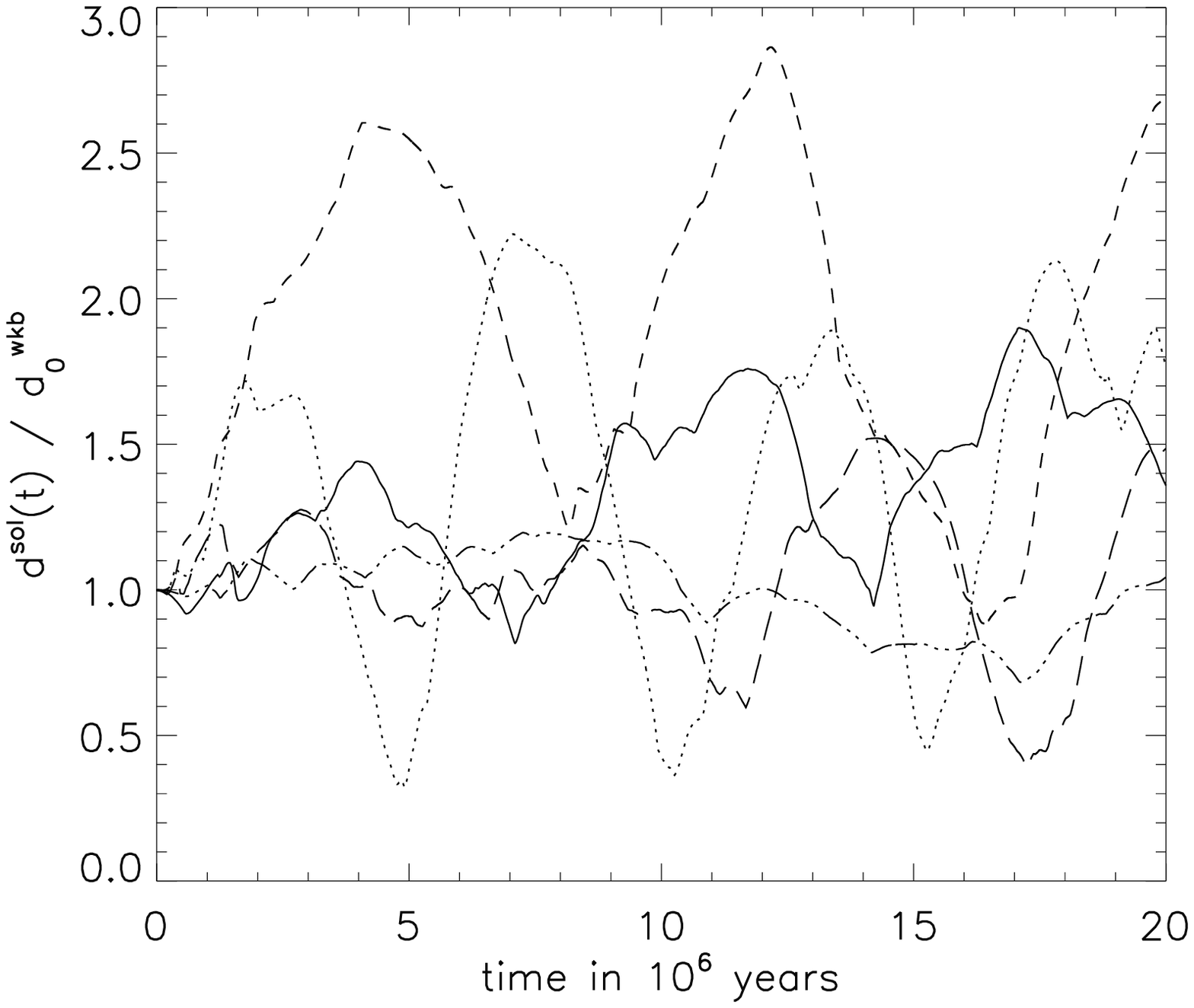}
           }
\caption{{\bf a)} The ratio of $\lambda_{\mathrm{A0}}$ to 
         the extension of the WKB solution determines how well
         $\langle d^{\mathrm{sol}} \rangle_{\mathrm{t}}$ and
         $d^{\mathrm{wkb}}_{\mathrm{0}}$ agree.  Shown is $\langle
         d^{\mathrm{sol}}\rangle _\mathrm{t} /
         d_{\mathrm{0}}^{\mathrm{wkb}}$ as a function of
         $\lambda_\mathrm{A0} / d_{\mathrm{0}}^{\mathrm{wkb}}$ for the
         different runs. The dotted line is the corresponding linear
         least square fit. {\bf b)} The spatial extension of the
         nonlinear solution is more or less time variable and can show
         distinct oscillations. For details on the oscillation period,
         see text. Shown is
         $d^{\mathrm{sol}}(t)/d^{\mathrm{wkb}}_{\mathrm{0}}$ as a
         functions of time (x-axis, in million years) for three
         oscillatory runs R20.20.25.4.10 (solid), R100.10.5.4.10
         (dashed), and R100.45.5.8.10 (dotted), as well as for two
         runs without a clear oscillation, R20.10.12.1.10
         (dash-three-dots) and R100.45.5.4.10 (long dashes).}
\label{fig:lamdep}
\end{figure*}
\subsection{Spatial extension of turbulent slab}
\label{sec:extension}
We find that the time averaged spatial extension of the solution
roughly agrees with the spatial extension of the corresponding WKB
solution. As a function of time, the spatial extension
$d^{\mathrm{sol}}(t)$ can oscillate but does not have to.
$d^{\mathrm{sol}}(t)$ we define in analogy with
$d_{\mathrm{0}}^{\mathrm{wkb}}$ (see Sect.~\ref{sec:dimensionless}) as
the distance where the column density reaches 90\% of ${\cal M}(t)$.
We similarly define the extension $d^{\mathrm{hs}}(t)$ and
$d^{\mathrm{wkb}}(t)$ of the corresponding hydrostatical and WKB
solutions, where corresponding means that at time $t$ the three
solutions have the same ${\cal M}(t)$. Finally, we denote by $\left <
 . \right >_{\mathrm{t}}$ the time average between $5\cdot 10^{6}$
and $2\cdot 10^{7}$ years.

\subsubsection{Dependence on system parameters}
\label{sec:extav}
In all our simulations, the time averaged extension $\langle
d^{\mathrm{sol}}\rangle _\mathrm{t}$ of the slab agrees to within
a factor of three with the spatial extension of the WKB solution
$d^{\mathrm{wkb}}_{\mathrm{0}}$. The dominant parameters governing
the extension of the slab are, therefore, the parameters governing the
WKB solution, i.e. $\rho_{0}$, $B_{\mathrm{\perp}}$, and $T$.

Deviations from the spatial extension of the WKB solution we find
to depend linearly on the dimensionless WKB parameter
$W_{\mathrm{\lambda A0}} = \lambda_\mathrm{A0} /
d^{\mathrm{wkb}}_{\mathrm{0}}$. Fig.~\ref{fig:lamdep}a shows the
ratio of the spatial extensions $\langle d^{\mathrm{sol}}\rangle
_\mathrm{t} / d^{\mathrm{wkb}}_{\mathrm{0}}$ as a function of
$\lambda_\mathrm{A0} / d^{\mathrm{wkb}}_{\mathrm{0}}$ for the
different runs. A linear least square fit, also shown in the figure,
gives a dependence $\langle d^{\mathrm{sol}} \rangle _\mathrm{t} =
\alpha \lambda_\mathrm{A0} + \beta d^{\mathrm{wkb}}_{\mathrm{0}}$ with
$\alpha = 4.06$ and $\beta = 0.36$. Linear fits of similar quality are
obtained if instead of $d^{\mathrm{wkb}}_{\mathrm{0}}$ one uses as
scaling parameter $\langle d^{\mathrm{hs}}\rangle _\mathrm{t}$, or
$\langle d^{\mathrm{wkb}}\rangle _\mathrm{t}$. For the last case, the
fitting parameters are $\alpha = 3.86$ and $\beta = 0.37$.  If instead
of $\lambda_\mathrm{A0}$ we consider the time average of the true
Alfv\'{e}n wave-length $ \langle \lambda_\mathrm{A} \rangle
_\mathrm{t} $ at or close to the slab center, we cannot identify any
such clear dependence.  $ \langle \lambda_\mathrm{A} \rangle
_\mathrm{t} $ is usually larger than $\lambda_\mathrm{A0}$ by a factor
of about 2 to 8.

From Fig.~\ref{fig:lamdep}a it can be seen that best agreement between
$\langle d^{\mathrm{sol}}\rangle _\mathrm{t}$ and
$d^{\mathrm{wkb}}_{\mathrm{0}}$ is obtained around
$\lambda_\mathrm{A0}/ d^{\mathrm{wkb}}_{\mathrm{0}} = 0.17$. The
deviations at larger values of $\lambda_\mathrm{A0} /
d^{\mathrm{wkb}}_{\mathrm{0}}$ are not too surprising. A basic
assumption for the validity of WKB theory is that the density changes
only on scales much larger than the Alfv\'{e}n wave-length. This
assumption fails to hold as $\lambda_\mathrm{A0} /
d^{\mathrm{wkb}}_{\mathrm{0}}$ increases. For the deviations at small
values of $\lambda_\mathrm{A0} / d^{\mathrm{wkb}}_{\mathrm{0}}$ we
have checked that they are not caused by a too coarse spatial
discretization, which would cause artificial wave-damping and thus
reduced support against self-gravity.  Increasing the spatial
resolution by a factor of four left the nonlinear solution
unchanged. Instead, the true reason again lies in the failure of the
WKB approach to be valid. The mass distribution in the system still
consists of thin, dense sheets separated by more diffuse medium. As
the sheet thickness is much less than the Alfv\'{e}n wavelength in the
tenuous medium, the conditions for a WKB description are not met. We
come back to this point in more detail in Sect.~\ref{sec:analext}.

Besides the dominant effect of $\lambda_\mathrm{A0}/
d^{\mathrm{wkb}}_{\mathrm{0}}$, Fig.~\ref{fig:lamdep}a suggests
$\beta_{\mathrm{0}}$ to have a second order effect.
Simulations with identical ratio $\lambda_{\mathrm{A0}}
/d^{\mathrm{wkb}}_{\mathrm{0}}$ usually have a larger extension for
smaller $\beta_{\mathrm{0}}$.

In view of what was said in Sect.~\ref{sec:dimensionless}, the average
slab thickness must equal the reference scale
$d_{\mathrm{0}}^{\mathrm{wkb}}$ multiplied by a function of
$\alpha_{\mathrm{0}}$, $\beta_{\mathrm{0}}$ and $W_{\mathrm{\lambda A
0}}$. In the limit of very small pressure the dependence on
$\beta_{\mathrm{0}}$ disappears. In the limit of very large
$W_{\mathrm{\lambda A 0}}$, that is very small $\lambda_{\mathrm{A0}}
/ d^{\mathrm{wkb}}_{\mathrm{0}}$, the dependence on
$W_{\mathrm{\lambda A 0}}$ also disappears. There may remain some
dependence on $\alpha_{\mathrm{0}}$, which Fig.~\ref{fig:lamdep}a
indicates to be weak. The ratio $\langle d^{\mathrm{sol}}
\rangle_{\mathrm{t}} / d^{\mathrm{wkb}}_{\mathrm{0}}$ then approaches
a value which is apparently close to $(1/2)$. For higher values of
$\lambda_{\mathrm{A0}} / d^{\mathrm{wkb}}_{\mathrm{0}}$, we expect
some dependence of $\langle d^{\mathrm{sol}} \rangle_{\mathrm{t}} /
d^{\mathrm{wkb}}_{\mathrm{0}}$ on this parameter, which may be
represented by a Taylor expansion to first order for not too large
$\lambda_{\mathrm{A0}} / d^{\mathrm{wkb}}_{\mathrm{0}}$. This is in
rough agreement with what is seen in Fig.~\ref{fig:lamdep}a, and is
consistent with the idea that the larger the Alfv\'{e}n wave-length,
the larger the spacing between dense sheets, and the thicker the
system.
\begin{figure*}[htp]
\centerline{
           }
\caption{For simulations with the same Alfv\'en wave-length 
  $\lambda_\mathrm{A0}$ and temperature $T$, but otherwise widely
  different parameters, the size of the voids (black, $N=10^{-1}$
  cm$^{-3}$) separating the high density sheets (white, $N=10^{4}$
  cm$^{-3}$) is about the same. Comparison with Fig.~\ref{fig:d3dtl}
  shows the effect of different values of $\lambda_\mathrm{A0}$ and
  $T$. The general density stratification leads to larger voids at
  larger distances. Shown is the logarithmic particle density
  as a function of space (x-axis, in units of 6 pc) and time (y-axis,
  in units of $10^{6}$ years) for runs {\bf a)} R20.20.25.4.10 and
  {\bf d)} R100.10.5.4.10.}
\label{fig:r20.20.25.4}
\end{figure*}
\subsubsection{Time variability of slab extension}
\label{sec:timevaria}
As a function of time, the spatial extension $d^{\mathrm{sol}}(t)$ of
the slab can be very variable but does not have to be.  We mention
already here that the substructure of the slab, voids and high density
sheets, always shows oscillatory motions (see Sect.~\ref{sec:densstru}
and Fig.~\ref{fig:r20.20.25.4}).  In Fig.~\ref{fig:lamdep}b, the ratio
of $d^{\mathrm{sol}}(t) / d^{\mathrm{wkb}}_{\mathrm{0}}$ is shown as a
function of time for five runs, representative of the total of our
simulations. As can be seen, only some of the runs show a strong
variability while others have a more or less constant extension.  A
measure for the strength of this time variability is the standard
deviation $\sigma_{\mathrm{d}}$ of $d^{\mathrm{sol}}(t) /
d^{\mathrm{wkb}}_{\mathrm{0}}$.  Using for the calculation of
$\sigma_{\mathrm{d}}$ the same time interval as for the time averages,
we find the relative standard deviation
$\sigma^{\mathrm{rel}}_{\mathrm{d}} =
\sigma_{\mathrm{d}}/(\langle d^{\mathrm{sol}} \rangle_{\mathrm{t}} /
d^{\mathrm{wkb}}_{\mathrm{0}})$ to lie in an interval of $ 0.1
\;\rlap{\lower 2.5pt \hbox{$\sim$}}\raise 1.5pt\hbox{$<$}\;
\sigma^{\mathrm{rel}}_{\mathrm{d}} 
\;\rlap{\lower 2.5pt \hbox{$\sim$}}\raise 1.5pt\hbox{$<$}\;
0.9$ (see
Table~\ref{tab:mod_params}).

In about half of our simulations, the variability takes the form of a
roughly periodic oscillation. The period of the oscillation is,
however, not always well defined.  Twice the sound crossing time of
the average slab thickness, $T_{\mathrm{s}} = 2 \langle
d^{\mathrm{sol}} \rangle_{\mathrm{t}} / c_{\mathrm{s}}$, is often
close to the observed period $T_{\mathrm{obs}}$.  The ratio
$T_{\mathrm{obs}}/T_{\mathrm{s}}$ averages over the different runs to
a value close to 0.95. However, this ratio, normalized to its average
value, has significant scatter of the order of 0.8. Moreover, the fact that
our slabs are more wave-supported than gas pressure supported makes it
difficult to understand a scaling like $T_{\mathrm{obs}} \sim
T_{\mathrm{s}}$ on physical grounds. A more natural scaling, as we
will show in Sect.~\ref{sec:oscanal}, would be that the oscillation
period is proportional to twice the crossing time of the average slab
thickness at velocity $v_{\mathrm{\perp 0}}$. For
$T_{\mathrm{v\perp 0}}=2
\langle d^{\mathrm{sol}} \rangle_{\mathrm{t}} / v_{\mathrm{\perp0}}$ 
the ratio $T_{\mathrm{obs}} \sim T_{\mathrm{v\perp 0}}$ averages, over
our different runs, to 3.4 with a scatter of 0.4.  For
$T_{\mathrm{v\perp 0}}=2 d^{\mathrm{wkb}}_{\mathrm{0}} /
v_{\mathrm{\perp0}}$ the average is 3.6 with a scatter of only 0.2.
Thus, to within a factor of order unity, but somewhat larger than
unity, our simulations are also consistent with $T_{\mathrm{obs}}$
being proportional to $T_{\mathrm{v\perp 0}}$.
\begin{figure*}[htp]
\centerline{
           }
\caption{Smaller voids (black, $N=10^{-1}$ cm$^{-3}$) separating 
         the high density sheets (white, $N=10^{4}$ cm$^{-3}$) are
         obtained for smaller $\lambda_{\mathrm{A0}}$ and / or higher
         temperatures $T$. {\bf a)} Run R20.20.12.4.10 has the same
         temperature $T$ as run R20.20.25.4.10 in
         Fig.~\ref{fig:r20.20.25.4}, but only half the value of
         $\lambda_{\mathrm{A0}}$.  {\bf b)} Run R20.20.25.4.40 has the
         same $\lambda_{\mathrm{A0}}$ as run R20.20.25.4.10 in
         Fig.~\ref{fig:r20.20.25.4} but a four times higher
         temperature $T$. Shown is the logarithmic particle density as
         a function of space (x-axis, in units of 6 pc) and time
         (y-axis, in units of $10^{6}$ years).}
\label{fig:d3dtl}
\end{figure*}
\subsection{Inhomogeneous structure of turbulent slab}
\label{sec:densstru}
Under the influence of both the injected wave and self-gravity, the
interior structure of the slab becomes very inhomogeneous,
substructure develops.  This substructure is the result of
time-dependent, nonlinear effects and as such is beyond the reach of
WKB theory. Again, we find $\lambda_{\mathrm{A0}}$ to play a critical
role, but now the temperature of the slab has an effect as well.

The density distribution is characterized by extended, low density
voids and narrow, high density sheets. In these low density voids, the
other variables, $\vec{v}$ and $\vec{B}$, remain nearly
constant. Therefore, the density substructure is a good mirror for
their substructure as well, and we restrict ourselves to density in
the following. To characterize the density substructure in our
gravitationally stratified slab, we represent the time series of 1D
density distributions as a 2D grey-scale plot. Low density regions
then appear as more or less extended (dark) patches, while high
density sheets take the form of thin (bright) lines.

Fig.~\ref{fig:r20.20.25.4} shows such a representation of the density
for two runs. Apparent also in this representation is the presence
(run R100.10.5.4.10) or absence (run R20.20.25.4.10) of a global
oscillation of the slab.  However, it can also be seen that,
independent of the existence of a global oscillation, the high density
sheets within the slab undergo oscillatory motions. The motion of a
single sheet is roughly parabolic, but may be interrupted
at any time as two sheets collide. Different sheets can have widely
different oscillation periods, which are mostly much smaller than the
period of the global oscillation of the slab.

This motion of the high density sheets, together with the number of
sheets, determines the size of the voids (dark patches in
Fig.~\ref{fig:r20.20.25.4}) in the 2D representation of the data. We
call this the scale of the substructure, a larger scale substructure
thus referring to an overall large size of the voids. Based on the
only qualitative measure of the grey-scale plots, we find that this
scale of the substructure is, in essence, determined by
$\lambda_{\mathrm{A0}}$ and temperature $T$.  The size of the
substructure in the two runs shown in Fig.~\ref{fig:r20.20.25.4},
which have the same $\lambda_{\mathrm{A0}}$ and temperature $T$, but
otherwise widely different parameters, is very similar.  This becomes
even more apparent if Fig.~\ref{fig:r20.20.25.4} is compared with
Fig.~\ref{fig:d3dtl}, where simulations with only half the Alfv\'{e}n
wave-length (Fig.~\ref{fig:d3dtl}a) and four times the slab
temperature $T$ (Fig.~\ref{fig:d3dtl}b) are shown. On the other hand,
the scale of the density distribution for the two simulations shown in
Fig.~\ref{fig:d3dtl} is again similar. This suggest that augmenting
the temperature by a factor of four has the same effect as reducing
the Alfv\'{e}n wave-length $\lambda_{\mathrm{A0}}$ by a factor of two.
Or, put otherwise, that the scale of the density substructure in space
and time depends on the ratio $ \sqrt{T} / \lambda_{\mathrm{A0}}$ or,
equivalently, on $\sqrt{\beta_{\mathrm{0}}}\omega$. We currently have,
however, no quantitative measure to further corroborate this
speculation.

The dominant role of $\lambda_{\mathrm{A0}}$ for the scale of the
density distribution may have been anticipated from what was said
towards the end of Sect.~\ref{sec:dimensionless} as well as from
parametric instability theory \citep{1978ApJ...224.1013D,
1978ApJ...219..700G}.  Analytical results for the linear regime
indeed show the separation of density disturbances to increase with
increasing Alfv\'{e}n wave-length.  On the other hand, the system we
consider is highly nonlinear and $\lambda_{\mathrm{A0}}$ soon loses
its meaning even on average.
\begin{figure}[htp]
\centerline{
\includegraphics[width=8.5cm]{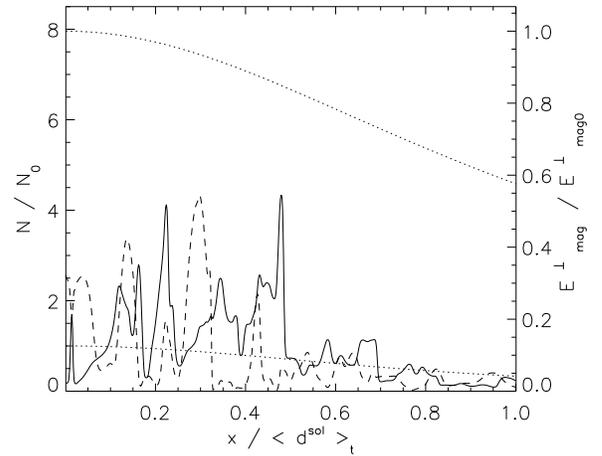}
           }
\caption{The energy $E^{\mathrm{\perp}}_{\mathrm{mag}}$  of the 
         transverse magnetic field is very inhomogeneous. It is
         generally smaller than in the corresponding WKB solution and
         not correlated with the density substructure.  Shown are the
         density (solid line) and the energy density of the transverse
         magnetic field (dashed line), normalized to their initial,
         central values, as functions of distance to the central plane
         (x-axis, in units $\langle d^{\mathrm{sol}}
         \rangle_{\mathrm{t}}$) at the example of simulation
         R20.20.5.4.10 at a time of $2.07 \cdot 10^{7}$ years.  The
         dotted curves denote $N/N_{\mathrm{0}}$ (lower curve) and
         $E^{\mathrm{\perp}}_{\mathrm{mag}} /
         E^{\mathrm{\perp}}_{\mathrm{mag0}}$ (upper curve) of the
         corresponding WKB solution.}
\label{fig:dens_emag}
\end{figure}
\begin{figure*}[htp]
\centerline{
\includegraphics[width=8.1cm,height=6.3cm]{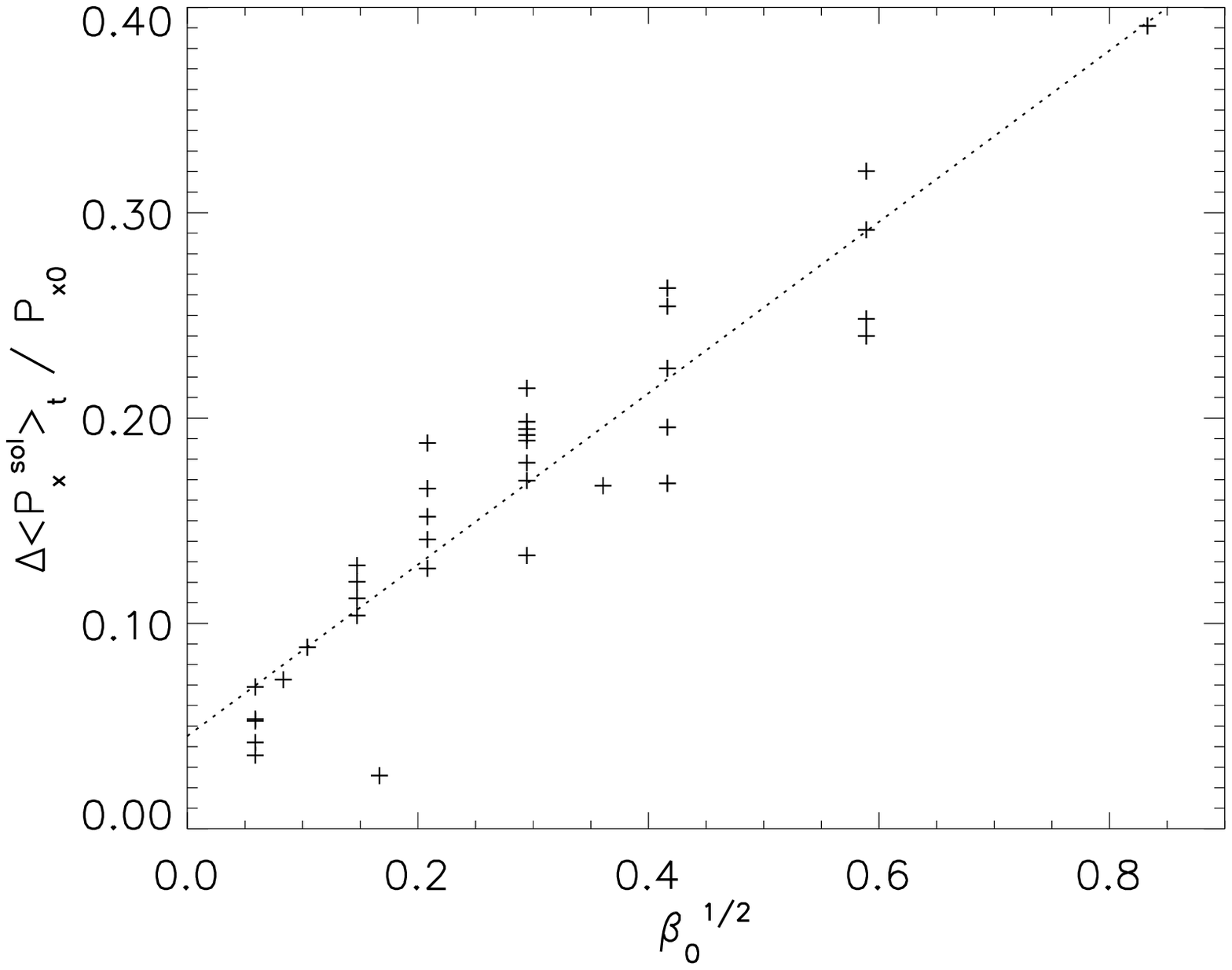}
\includegraphics[width=8.1cm,height=6.3cm]{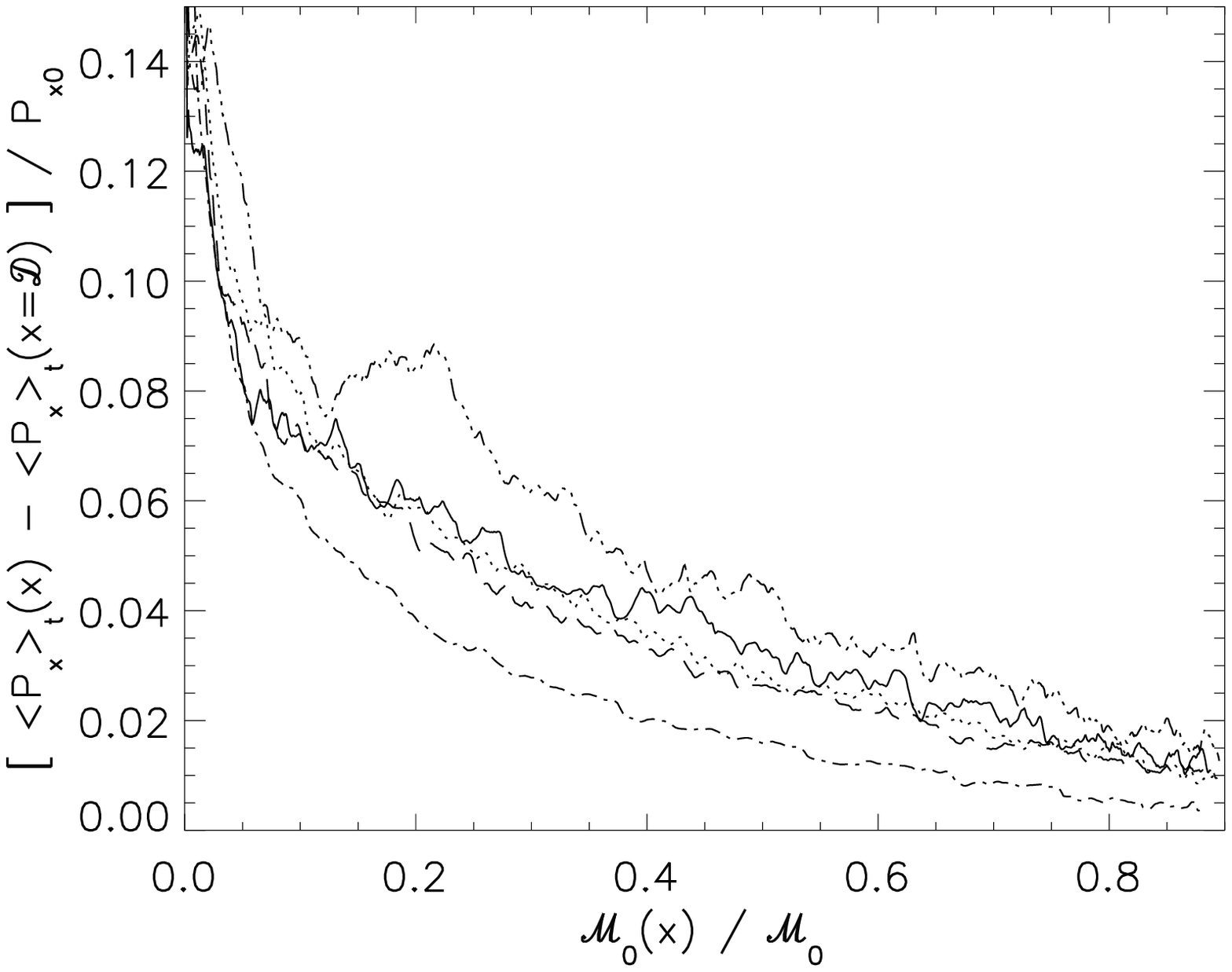}
           }
\caption{ {\bf a)}  Change of time averaged Poynting-flux, between the
  central plane and $\langle d^{\mathrm{sol}} \rangle_{\mathrm{t}}$,
  in units of the corresponding WKB Poynting flux, $P_{\mathrm{x0}}$
  as a function of $\sqrt{\beta_{\mathrm{0}}}$ for the different runs
  we have performed.  {\bf b)} Change of time averaged Poynting-flux,
  $\Delta \langle P^{\mathrm{sol}}_{\mathrm{x}} \rangle _\mathrm{t}$,
  in units of the corresponding WKB Poynting flux, as a function of
  increasing, relative column density ${\cal M}_{\mathrm{0}}(x)/{\cal
  M}_{\mathrm{0}}$ for five runs, all with $\beta_{\mathrm{0}} =
  0.087$: R20.20.5.4.10 (dash-dotted), R20.20.10.4.10 (long dashes),
  R20.20.12.4.10 (dotted), R20.20.17.4.10 (solid), R20.20.25.4.10
  (dash-three-dots). }
\label{fig:dpfx}
\end{figure*}
\subsection{Energy of turbulent slab}
At the central plane, we constantly feed energy into the slab. The
energy provided by this 'one point forcing' penetrates far into the
slab, despite the gradual destruction through isothermal shocks and
viscous dissipation.  Note that the viscous dissipation is merely
given by the numerical method and the discretization, and may not
accurately mimic real viscosity, diffusion, or resistivity.  The
energy density of the transverse magnetic field is far from smooth and
is generally smaller than its initial WKB value. A typical situation
at later times is shown in Fig.~\ref{fig:dens_emag}. The density
substructure leads to wave reflection and shock
formation. Occasionally, the WKB value of
$E^{\mathrm{\perp}}_{\mathrm{mag}}$ can be exceeded, as two high
density sheets move against each other, thus compressing the
transverse magnetic field between them.
\subsubsection{Loss of Poynting-flux}
Unlike in the WKB solution, the Poynting-flux in our numerical
solution is neither constant in space nor time.
Let us denote by $P_{\mathrm{x0}}$ the Poynting-flux of the initial 
WKB solution and by $\langle P_{\mathrm{x}} \rangle
_\mathrm{t}$ the time averaged
x-component of the true Poynting-flux. 
We find that the change
$\Delta \langle P^{\mathrm{sol}}_{\mathrm{x}} \rangle _\mathrm{t}$ of the
Poynting-flux over a distance $\langle d^{\mathrm{sol}}
\rangle_{\mathrm{t}}$ from the central plan of the slab lies 
in a range $ 0.03 \;\rlap{\lower 2.5pt \hbox{$\sim$}}\raise
1.5pt\hbox{$<$}\; \Delta \langle P^{\mathrm{sol}}_{\mathrm{x}} \rangle
_\mathrm{t}/P_{\mathrm{x0}}
\;\rlap{\lower 2.5pt \hbox{$\sim$}}\raise 1.5pt\hbox{$<$}\; 0.4$. The
ratio depends about linearly on $\sqrt{\beta_{\mathrm{0}}}$, as can be
seen in Fig.~\ref{fig:dpfx}a.  In view of the results in
Sect.~\ref{sec:densstru}, this dependence suggests a connection
between the loss of Poynting-flux and the amount of density structure
we have in the slab. There we have seen that larger values of
$\sqrt{T}/\lambda_{\mathrm{A0}}$, which implies larger values of
$\sqrt{\beta_{\mathrm{0}}}$, correlate with a finer scale density
structure.

Largest losses of Poynting-flux occur in the innermost region of the
slab, and more than 90\% of the total loss of Poynting-flux over the
entire computational domain occurs within a distance $\langle
d^{\mathrm{sol}} \rangle_{\mathrm{t}}$ of the central plane of the
slab.  Fig.~\ref{fig:dpfx}b shows the spatial variation of the loss of
Poynting-flux. There, $\langle P_{\mathrm{x}} \rangle _\mathrm{t}(x) -
\langle P_{\mathrm{x}} \rangle _\mathrm{t}(x={\cal D})$ is shown, in units of
$P_{\mathrm{x0}}$, as a function of increasing, relative column
density ${\cal M}_{\mathrm{0}}(x)/{\cal M}_{\mathrm{0}}$ for five
runs. Except for the value of $\omega$, the runs have identical
parameters with a value of $\beta_{\mathrm{0}}=0.087$.  As can be
seen, the decrease of Poynting-flux in this representation is fairly
similar for all five simulations. Note, however, that the actual
spatial extension of the five runs is very different ($\langle
d^{\mathrm{sol}} \rangle_{\mathrm{t}} = 0.11$\,${\cal D}$ for
R20.20.5.4.10 but $\langle d^{\mathrm{sol}} \rangle_{\mathrm{t}} =
0.34$\,${\cal D}$ for R20.20.25.4.10).

The change of Poynting-flux is part of the change of the total energy
flux.  In fact, we find $\Delta
\langle P^{sol}_{\mathrm{x}} \rangle _\mathrm{t}$ to amount to between
60\% and 90\% of the change of the total energy flux between the
central plane of the slab and $\langle d^{\mathrm{sol}}
\rangle_{\mathrm{t}}$. The remaining amount is made up from the fluxes
of gravitational and kinetic energy, the latter usually contributing
considerably more. As the slab is, in time average, in a
quasi-stationary state, this change in energy flux between the central
plane and $\langle d^{\mathrm{sol}}
\rangle _\mathrm{t}$ must be radiated.
\subsubsection{Energy equipartition}
In all our simulations, the ratio of the time averaged energy density
of the transverse magnetic field and of the energy density associated
with the transverse velocity, $\langle
E^{\mathrm{\perp}}_{\mathrm{mag}}
\rangle_{\mathrm{t}} / \langle E^{\mathrm{\perp}}_{\mathrm{kin}}
\rangle_{\mathrm{t}}$, shows some spatial variation. 
Within about the innermost 10\% of $\langle d^{\mathrm{sol}}
\rangle_{\mathrm{t}}$, we find for $\langle
E^{\mathrm{\perp}}_{\mathrm{mag}} \rangle_{\mathrm{t}} /
\langle E^{\mathrm{\perp}}_{\mathrm{kin}} \rangle_{\mathrm{t}}$
typical values of about 2, for a few simulations even somewhat higher.
Further out, to distances around $\langle
d^{\mathrm{sol}}\rangle_{\mathrm{t}}$, we observe approximate
equipartition between $\langle
E^{\mathrm{\perp}}_{\mathrm{mag}}
\rangle_{\mathrm{t}}$ and $\langle E^{\mathrm{\perp}}_{\mathrm{kin}}
\rangle_{\mathrm{t}}$. The time averaged 
kinetic and magnetic energy densities themselves decrease with
increasing distance from the central plane of the slab.

If we look at the time and space averaged energy quantities, 
$\langle E^{\mathrm{\perp}}_{\mathrm{mag}}
\rangle_{\mathrm{t,x}}$ and $\langle E^{\mathrm{\perp}}_{\mathrm{kin}}
\rangle_{\mathrm{t,x}}$, we find approximate equipartition in most 
of our simulations.  Here, $\langle \cdot \rangle_{\mathrm{t,x}}$
denotes time average and subsequent spatial average, the later taken
over $x \in [0,\langle d^{\mathrm{sol}} \rangle_{\mathrm{t}}]$.
$\langle E^{\mathrm{\perp}}_{\mathrm{mag}} \rangle_{\mathrm{t,x}} /
\langle E^{\mathrm{\perp}}_{\mathrm{kin}} \rangle_{\mathrm{t,x}}$ lies
in a range between 0.8 and 1.2 except for five runs (R100.45.5.8.10,
R20.20.5.1.10, R20.10.2.1.10, R20.20.5.1.20 with a ratio between 0.6
and 0.8 and R100.45.1.4.10 with a ratio of 0.4).  For the time
and space average of the total kinetic energy density $\langle
E_{\mathrm{kin}}\rangle_{\mathrm{t,x}}$, we find the ratio $\langle
E^{\mathrm{\perp}}_{\mathrm{kin}}\rangle_{\mathrm{t,x}} / \langle
E_{\mathrm{kin}} \rangle_{\mathrm{t,x}}$ to lie in a range between 0.7
and 0.9, except for run R100.20.5.4.10 where this ratio is only 0.6.
\section{Discussion}
\label{sec:discussion}
\subsection{Spatial extension of a non-homogeneous slab}
\label{sec:analext}
In Sect.~\ref{sec:num_results} we have seen that the WKB extension
provides only a first approximation to the spatial extension of the
fragmented slab. In the following, we give a qualitative explanation
of why the extension of a fragmented slab differs from its WKB
extension, in particular also for small values of
$\lambda_{\mathrm{A0}} / d^{\mathrm{wkb}}_{\mathrm{0}}$, the seeming
WKB limit. The WKB picture considers a diffuse mass distribution. The
thickness of the layer of material between column density 0 (the
central plane) and column density ${\cal M}$ is
\begin{equation}
H({\cal M}) = \int_{0}^{\cal M} \frac{d{\cal M}'}{\rho({\cal M}')}
\label{eq:hm}
\end{equation}
where $\rho({\cal M})$ is the density of the matter sitting on top of
an amount of material of total column density ${\cal M}$. If a slab is
fragmented and contains dense sheets separated by tenuous regions, the
contribution of the sheets to the integral in Eq.~\ref{eq:hm} is very
small. Thus the system's thickness is essentially due to the tenuous
intersheet regions.  This indicates that an entirely diffuse slab
would be thicker than one which has a substantial fraction of its mass
in the form of dense sheets, provided that the tenuous regions of the
diffuse slab have drastically smaller density than the intersheet
regions of the fragmented slab.

In a limit where the WKB approximation is close to being valid, however,
$\rho({\cal M})$ should be quite similar in both cases. To see this, let us
denote by $B_{\perp}({\cal M})$ the magnetic amplitude of the wave at a point
sitting on top of an amount of material of total column density
${\cal M}$. Suppose the slab contains a large number of low-column-density
thin sheets levitating in equilibrium under the pressure of the
Alfv\'enic flux.  Assuming equilibrium yields a relation between
the jump $dB_{\perp}$ of $B_{\perp}$ at the crossing of one such sheet
and the corresponding jump in column density $d{\cal M}$: 
\begin{equation}
d \left(\frac{B_{\mathrm{\perp}}^2}{4 \mu_{\mathrm{0}}} \right )= 
- 4 \pi G {\cal M} \mbox{\,\,} d{\cal M}.
\end{equation}
The same relation also holds for any infinitesimal slab of
tenuous material in equilibrium. Thus, irrespective of whether the
matter has an entirely diffuse distribution or is partially fragmented
in a large number of thin sheets, the same relation between
$B_{\perp}^2({\cal M})$ and ${\cal M}$ holds. This means that the distribution of
wave support as a function of column density is independent of the
details of the density distribution. If the intersheet medium can also
be described by a WKB approach, it would result that the density
distribution $\rho({\cal M})$ would be the same in the entirely tenuous slab
as it is in the tenuous regions of the fragmented slab. A glance at
Eq.~\ref{eq:hm} then shows that the fragmented slab would be less
extended, by a factor which depends on which fraction of its mass is
in the form of dense sheets.
\subsection{Oscillation of slab extension}
\label{sec:oscanal}
In Sect.~\ref{sec:timevaria} we have seen that the spatial extension
of the slab can oscillate on a time scale of a few Myr. That a period
on the order of $T_{\mathrm{v\perp0}} = 2 \langle d^{\mathrm{sol}}
\rangle_{\mathrm{t}}/ v_{\mathrm{\perp0}}$ is to be expected on physical 
grounds shows the following analysis.

We discuss the fundamental period of deviations from equilibrium of an
isothermal slab, supported against self-gravity by gas pressure and 
a flux of WKB Alfv\'en waves. Such a model is, admittedly, far from
our highly structured slabs, but it should be
sufficient to understand the main parameters which control the global
oscillation. In this model, the force exerted by Alfv\'{e}n waves
takes the form of the gradient of the wave energy density $U$. The
wave energy flux $ U c_{\mathrm{A\parallel}}$ we assume to be constant
in space at any time, with $c_{\mathrm{A\parallel}}$ the Alfv\'{e}n velocity
associated with $B_{\mathrm{\parallel}}$ and the local mass density.
The equilibrium is described by functions $\rho_{\mathrm{e}}(x)$,
$U_{\mathrm{e}}(x)$ and ${\cal M}_{\mathrm{e}}(x)$, the latter being the
column density between 0 and $x$. An additional subscript 0 denotes
the values of these functions at $ x = 0$.  In particular,
$U_{\mathrm{e0}} =
\rho_{\mathrm{e0}} v_{\mathrm{e0\perp}}^2 /2$.  As a boundary
condition for the perturbation, we impose, in analogy with our
simulations, that the velocity amplitude of the wave injected at $x=0$
is fixed in time. This has the important consequence that the wave flux
forced into the slab at the central plane is modulated by the varying
gas density there. Indeed, for fixed $B_{\mathrm{\parallel}}$ and
$v_{\mathrm{\perp}}$ the sum of the kinetic and Poynting Alfv\'en wave
energy flux, $P = \rho v_{\mathrm{\perp}}^2
c_{\mathrm{A\parallel}}/2$, is proportional to $\sqrt{\rho}$.

\begin{figure}[tp]
\centerline{
\includegraphics[width=8.5cm]{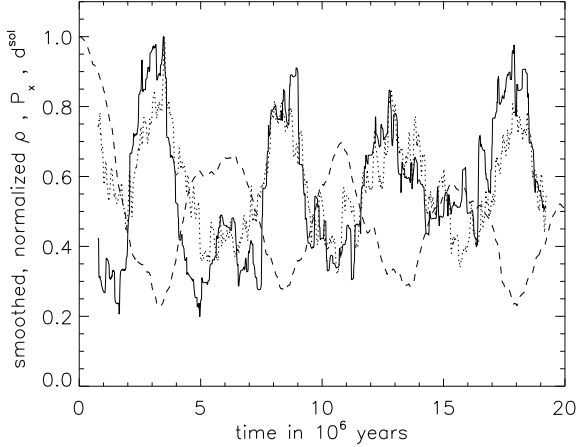}
           }
\caption{The global oscillation of the slab, if existent, 
is in antiphase with the modulation of the running mean in time (time
window 1.5 Myr) of the Poynting-flux and the density at the central
plane of the slab. Shown are, as a function of time in $10^{6}$ years,
the normalized spatial extension of the slab (dashed line), and the
normalized running mean in time of the Poynting-flux (dotted line) and
of the density (solid line). Normalized means that the quantities were
divided by the maximum value they assume in the time interval shown.}
\label{fig:rhopfxdsol}
\end{figure}

In about half of our simulations, we indeed observe such a modulation
of the Poynting-flux and density close to the central plane of the
slab, after applying a running mean in time to filter out the strong
short term variability stemming from individual high density sheets
(see Fig.~\ref{fig:rhopfxdsol}). The modulation of the Poynting-flux
then is in phase with the density perturbation and in antiphase with
the slab thickness, as it should be.

Linearizing the equations of motion about the equilibrium and assuming
a monochromatic perturbation, ${\cal M}_{\mathrm{1}}(x) e^{-i
\omega t}$ of the column density, we find that 
${\cal M}_{\mathrm{1}}(x)$ is a solution of the following linear
homogeneous equation:
\begin{eqnarray}
& & \rho_{\mathrm{e}} c_{\mathrm{s}}^2 
\frac{d}{dx}
\left( \frac{1}{\rho_{\mathrm{e}}} \frac{d {\cal M}_{\mathrm{1}}}{dx} \right) +
\frac{U_{\mathrm{e}}}{\rho_{\mathrm{e}}} \rho_{\mathrm{e}}^{3/2}
\frac{d}{dx} \left( \frac{1}{\rho_{\mathrm{e}}^{3/2}}
\frac{d {\cal M}_{\mathrm{1}}}{dx} \right) +  \nonumber \\
& & (\omega^2 + 4 \pi G \rho_{\mathrm{e}})
{\cal M}_{\mathrm{1}} + 
\frac{{\cal M}'_{\mathrm{1}}(0)}{ 2 \rho_{\mathrm{e}}}
\frac{dU_{\mathrm{e}}}{dx} = 0.
\label{eq:modeq}
\end{eqnarray}
The last term in Eq.~\ref{eq:modeq} represents the modulation in time
of the Poynting-flux entering the slab at $x = 0$. It is proportional
to the mass density perturbation at this point, ${\cal
M}'_{\mathrm{1}}(0) = d{\cal M} /dx = \rho_{\mathrm{1}}(0)$.  Had we
assumed that the slab suffers perturbations under a constant wave
energy flux, this term would have been absent from Eq.~\ref{eq:modeq}.
The spectrum associated with Eq.~\ref{eq:modeq} may only be found
numerically, but its gross features may be anticipated from a simple
dimensional analysis, substituting $(ik{\cal M}_{\mathrm{1}})$ for
$(d{\cal M}_{\mathrm{1}} /dx)$ and $(- U_{\mathrm{0}}/H_{\mathrm{e}})$ for
$(dU_{\mathrm{0}}/dx)$. $H_{\mathrm{e}}$ is the thickness of the
equilibrium slab, approximately given by the relation
\begin{equation}
4 \pi G
\rho_{\mathrm{e0}}^2 H_{\mathrm{e}}^2 = (U_{\mathrm{e0}} +
\rho_{\mathrm{e0}} c_{\mathrm{s}}^2).  
\label{eq:eqrel}
\end{equation}
This rough procedure does not make justice of the specific profile of
$\rho_{\mathrm{e}}(x)$. To take care of this, we introduce numerical
coefficients, $\alpha_{\mathrm{1}}$, $\alpha_{\mathrm{2}}$,
$\alpha_{\mathrm{3}}$, $\alpha_{\mathrm{4}}$, all positive and of
order unity. From Eq.~\ref{eq:modeq}, the dispersion relation is
expected to be of the form:
\begin{equation}
\omega^2 =
\alpha_{\mathrm{1}} \ c_{\mathrm{s}}^2 k^2 + 
\alpha_{\mathrm{2}} \ \frac{v_{\mathrm{\perp} e0}^2}{2} k^2 -
\alpha_{\mathrm{3}} \ 4 \pi G \rho_{\mathrm{e0}} + 
\alpha_{\mathrm{4}} \frac{i k v_{\mathrm{\perp} e0}^2}{2 H_{\mathrm{e}}}.
\label{eq:disprel}
\end{equation}
The last, imaginary, term of Eq.~\ref{eq:disprel} represents the last
term of Eq.~\ref{eq:modeq}, associated with the modulation of the wave
energy flux.  For the fundamental mode, $k \approx
H_{\mathrm{e}}^{-1}$. Its frequency should then be given by an 
expression of the form:
\begin{equation}
\omega^2 = 
4 \pi G \rho_{\mathrm{e0}} 
\left( \alpha_{\mathrm{1}} \ 
\frac{c_{\mathrm{s}}^2}{c^2} + 
\alpha_{\mathrm{2}} \  
\frac{ v_{\mathrm{\perp}e0}^2/2 }{ c^2} - 
\alpha_{\mathrm{3}} + i \alpha_{\mathrm{4}} \ 
\frac{ v_{\mathrm{\perp}e0}^2/2 }{ c^2}
\right),
\label{eq:fundmond}
\end{equation}
where $c^{2}=c_{\mathrm{s}}^2 + v_{\mathrm{\perp}e0}^2/2$.  This shows
that the fundamental time must be, to within a factor of order unity
but presumably smaller due to the presence of the negative
term $- \alpha_{\mathrm{3}}$, the Jeans frequency $(4 \pi G
\rho_{\mathrm{e0}})^{1/2}$. From Eq.~\ref{eq:eqrel}, the associated time  is,
for $c_{\mathrm{s}} << v_{\mathrm{e0\perp}}$, of order of
$\sqrt{2}H_{\mathrm{e}}/v_{\mathrm{e0\perp}}$.  Given the uncertainty
left by this analysis on the unknown factors
$\alpha_{\mathrm{1}}$ - $\alpha_{\mathrm{4}}$, this is in rough
agreement with the results of our simulations. The
modulation of the Alfv\'en flux injected in the slab, caused
by the modulation of the central density perturbation,
$\rho_{\mathrm{1}}(0)$, constitutes a feedback which endows $\omega^2$
with an imaginary part, and thus induces instability, even when the
system is stable under constant Alfv\'en wave energy flux.  It is
interesting to note that such a modulation is indeed present in our
simulations. The coherent oscillations observed in our calculations
may represent a nonlinear state of development of such an
instability.  At high amplitude, the oscillations of the different
fluid elements would cease to be isochronic and their initial
coherence may be lost.
\begin{figure*}[htp]
\centerline{
           }
\caption{Good enough spatial resolution is crucial to observe
         the structuring and supporting effect of the injected
         Alfv\'{e}n-wave. Shown is the same simulation,
         R20.20.5.4.10, once, {\bf a)}, with our usual spatial
         resolution of 5000 cells and once, {\bf b)}, with a reduced
         resolution of only 1000 cells. Logarithmic particle density
         is shown as a function of space (x-axis, in units of 6 pc)
         and time (y-axis, in units of $10^{6}$ years). A similar loss
         of structuring and support we observe if the order of the
         integration scheme is reduced.}
\label{fig:spaceres}
\end{figure*}
\subsection{Driving of turbulence and size of structure}
The one-point forcing we apply in our simulations is in contrast to
most other investigation of highly compressible turbulence, where the
forcing is applied at each grid point (see
e.g. \citet{maclow:99} for a description of such forcing and
\citet{gammie-ostriker:96} for the particular case of a 1D-slab).
Our results show that also such one point forcing is perfectly capable
of supporting and structuring a slab over large distances. The
injected waves are very efficient in distributing the driving energy
over a large spatial range. 

For the distance over which the injected energy is spread, as well as
for the scale of the density structure that forms, the driving
wave-length $\lambda_{\mathrm{A0}}$ is decisive.  This is not too
surprising. In Sect.~\ref{sec:dimensionless} we have seen that there
is no other small length scale likely to play a role in the problem we
consider. It then appears quite natural that $\lambda_{\mathrm{A0}}$
governs the structure size. Remembering that smaller scale structure
means more inhomogeneity on a smaller spatial scale, a faster
destruction of the injected energy for smaller $\lambda_{\mathrm{A0}}$
also seems natural.

That local energy injection can be sufficient to drive turbulence in a
whole volume has also been reported for the case of the interaction
zone of hypersonically colliding flows in 2D plane-parallel
hydrodynamics~\citep{1998A&A...330L..21W, 2000Ap&SS.274..343W}.
Although energy there is injected only at the oblique shocks confining
the interaction zone, supersonic turbulence persists in the whole
volume. A correlation between driving
wave-length and structure size exists as well (Folini \& Walder, in
preparation).  A correlation between structures  size and the
wave-length of the driving was also reported by \citet{maclow:99} for
hydrodynamic turbulence in a 3D periodic box, which was
monochromatically forced at each grid point.
\subsection{Numerical resolution}
\label{sec:disc_num}
We have experimented with different spatial resolutions and different
orders of integration for the MHD equations. Not surprising, we find
that simulations with shorter wave-lengths of the Alfv\'en waves are
more delicate in terms of spatial resolution and order of integration.
For run R20.20.5.4.10 we have verified that our results remain
essentially unchanged (mean quantities differ by a few percent at
most) if we further increase the spatial resolution. For coarser
discretizations (using 1000 cells instead of 5000) or lower orders
of integration, strong damping of the Alfv\'en wave due to numerical
diffusion occurs and both the density substructure as well as the
additional support against self-gravity are essentially lost, as can
be seen in Fig.~\ref{fig:spaceres}. For run R20.20.5.4.10, for
example, $\langle d^{\mathrm{sol}}/d_{\mathrm{0}}^{\mathrm{wkb}}
\rangle_{\mathrm{t}}$ is reduced from 0.46 to 0.32, for run
R20.20.10.4.10 the value is reduced from 0.75 to 0.48.  With regard to
numerical simulations, these results show that the problem we
investigated in this paper is currently out of reach for 3D
simulations.
\subsection{Connection to high density cores}
We, finally, would like to return to the larger frame, molecular
clouds and the formation of dense cores.  As mentioned earlier, both
theoretical and observational results today indicate that the
immediate vicinity of a forming dense core is highly structured. Most
striking are the observations by~\citet{2001ApJ...555..178F}
of the filaments around the starless dense core L1512. Various
explanations for their existence, all employing magnetic fields, seem
plausible~\citep{2000MNRAS.311...85F, 2000A&A...357.1143F,
del-zanna-et-al:01}. It seems most likely that such structures, or
rather the responsible physical processes, affect the formation and
accretion rate of dense cores.

Detailed studies of this structuring and its consequences are
mandatory to better link turbulence in molecular clouds with star
formation. Such studies are, however, only at their beginning. The 1D
slab we studied here is by no means an accurate model of a molecular
cloud. Nevertheless, it shows that magnetic waves can play a crucial
role in counteracting self-gravity. The 3D grid studies
by~\citet{2001ApJ...547..280H}, mentioned in the introduction, point
in the same direction. And while the 2D and 3D MHD simulations
by~\citet{del-zanna-et-al:01} of a plan parallel slab focus on
structure formation rather than on accretion, they show that density
filaments form, more or less aligned with the background magnetic
field.

With regard to future simulations in this field, existing simulations
show spatial resolution to be decisive.  Another crucial issue is
dimensionality. The 'fingering' observed by~\citet{del-zanna-et-al:01}
- and by~\citet{2001ApJ...555..178F} - clearly cannot be studied in a
1D model. And, as the authors showed as well, it can only be observed
if open boundaries are used, not periodic ones. Finally, and probably
most crucial of all, a future model of dense core formation will have
to account for the generation of magnetic waves
self-consistently. This may be the strongest reason to aim, in a next
step, at multidimensional models.
\section{Conclusions}
\label{sec:conclusions}
For the case of a 1D plane-parallel, self-gravitating slab we have
shown by means of numerical simulations and dimensional analysis the
following:

1) One source of energy injection is sufficient to sustain turbulence
throughout the slab.

2) The time-averaged spatial extension of the turbulent slab is
comparable to the extension of the corresponding WKB
solution. Deviations are at most a factor of three and depend about
linearly on the ratio of the initial central Alfv\'{e}n wave-length
and the extension of the WKB solution.

3) The scale of the substructure is governed by the initial central
Alfv\'{e}n wave-length and the temperature. Larger wave-lengths and
smaller temperatures lead to larger scale structure.

4) The energy loss, and thus the energy radiated by the slab, is
dominated by the loss of Poynting-flux, which increases almost linearly 
with $\sqrt{\beta_{\mathrm{0}}}$.

5) Within the slab, the energy density of the transverse magnetic field 
and the kinetic energy density of the transverse velocities are
in approximate equilibrium. The latter accounts for between 
70\% and 90\% of the total kinetic energy density.

(6) A too coarse mesh or a low order scheme leads to substantial
wave-damping, thus to loss of structuring and support.
\begin{acknowledgements}
D.F. was supported by a grant of the Swiss National Science
Foundation, grant number 8220-056553. R.W. was supported by the
Scientific Discovery through Advanced Computing (SciDAC) program of
the US-DOE, grant number DE-FC02-01ER41184. R.W. thanks the Institut
f\"ur Astronomie at ETH Z\"urich for its hospitality and providing a
part-time office. J.H. thanks the EC Platon program HTRN-CT-2000-00153
and the Platon group for helpful discussions.  The 
calculations were done on the NEC SX5 at IDRIS, Paris,
France, and on the CRAY SV1B at ETH Z\"urich, Switzerland. The authors
would like to thank the referee Dr. Hanawa for his valuable comments
which helped to improve the paper.
\end{acknowledgements}
\appendix
\section{Simulation parameters}
\begin{table*}
\caption{
The parameters used in the different simulations.
The first column contains the name of each run. 
These names reflect the free parameters of each run.
For example, R20.10.25.4.40 is the simulation with
$B_{\mathrm{\parallel}}=20\mu$G, $B_{\mathrm{\perp}}=10\mu$G,
$2\pi/\omega= 25 \cdot 10^{4}$ years, a central density of $N_{\mathrm{0}} = 4 \cdot
250$ particles per cm$^{3}$, and $T=40$K.
Subsequent
columns, from left to right, denote: 
$\lambda_\mathrm{A0} = c_\mathrm{A0} 2 \pi / \omega$: initial central
Alfv\'{e}n wave-length in $2.176 \cdot 10^{17}$ cm;
${\cal M}_\mathrm{0}/(m_\mathrm{H}{\cal D})$: initial mean particle
density in cm$^{-3}$;
${\cal M}_{\mathrm{0}}/{\cal M}^{\mathrm{\infty}}_{\mathrm{0}}$: ratio
of column densities within computational domain and in infinite space
of initial WKB solution;
$d^{\mathrm{wkb}}_{\mathrm{0}}/d^{\mathrm{hs}}_{\mathrm{0}}$: ratio of
extensions of WKB and hydrostatic solution at $t=0$;
$d^{\mathrm{wkb}}_{\mathrm{0}}/{\cal D}$: spatial extension of initial
WKB solution in units of computational domain;
$\alpha_{\mathrm{0}} = B_{\mathrm{\perp}}^{2} / B_{\mathrm{\parallel}}^{2}$: dimensionless parameter;
$\beta_{\mathrm{0}} = 2 c_{\mathrm{s}}^{2} / c_\mathrm{A0}^{2}$: central, initial  plasma beta; 
$W_{\mathrm{0}} = \omega / \sqrt{2 \pi G
\rho_{\mathrm{0}}}$: WKB parameter; 
$W_{\mathrm{\lambda A0}} = \lambda_\mathrm{A0}/d^{\mathrm{wkb}}_{\mathrm{0}}$: 
alternative WKB parameter;
$\langle d^{\mathrm{sol}} \rangle_{t}/{\cal D}$: 
time averaged spatial extension of solution, average taken between $5
\cdot 10^{6}$ years and $2 \cdot 10^{7}$ years, in units of total
domain ${\cal D}$;
$\langle d^{\mathrm{sol}} \rangle_{t}/d^{\mathrm{wkb}}_{\mathrm{0}}$: 
ratio of time averaged spatial extension of solution and extension of
corresponding, initial WKB solution;
$\sigma_{\mathrm{d}}^{\mathrm{rel}}$: relative standard deviation of
$d^{\mathrm{sol}}(t)/d^{\mathrm{wkb}}_{\mathrm{0}}$; 
${\cal M}_\mathrm{20}/{\cal M}_\mathrm{0}$: ratio of the column
densities at times $t=20 \cdot 10^{6}$ years and $t=0$;
$\Delta P$: time averaged, percental change of Poynting-flux between
central plane and $\langle d^{\mathrm{sol}} \rangle_{\mathrm{t}}$ in
units of the initial WKB Poynting-flux, $\Delta P = 100 \cdot {\langle
\Delta P^{\mathrm{sol}}_{\mathrm{x}} \rangle_{t}} / P_{\mathrm{x0}}$.}
\begin{center}
\begin{tabular}{|l||c|c||c|c|c|c|c|c|c||c|c|c|c|c|} 
\hline 
\arrayrulewidth 6.5cm
\rule[-0.35cm]{0.cm}{0.85cm}
run                                                                              &
$\lambda_\mathrm{A0}$                                                            &
$\frac{{\cal M}_\mathrm{0}}{m_\mathrm{H}{\cal D}}$                               & 
$\frac{{\cal M}_{\mathrm{0}}}{{\cal M}^{\mathrm{\infty}}_{\mathrm{0}}}$          &
$\frac{{d}^{\mathrm{wkb}}_{\mathrm{0}}}{{d}^{\mathrm{hs}}_{\mathrm{0}}}$         & 
$\frac{{d}^{\mathrm{wkb}}_{\mathrm{0}}}{\cal D}$                                 &
$\alpha_{\mathrm{0}}$                                                            &
$\beta_{\mathrm{0}}$                                                             &
$W_{\mathrm{0}}$                                                                 & 
$W_{\mathrm{\lambda A 0}}$                                                       & 
$\frac{\langle {d}^{\mathrm{sol}} \rangle_{t} }{{\cal D}}$                       &
$\frac{\langle {d}^{\mathrm{sol}} \rangle_{t}}{{d}^{\mathrm{wkb}}_{\mathrm{0}}}$ &
$\sigma_{\mathrm{d}}^{\mathrm{rel}}$                                             & 
$\frac{{\cal M}_\mathrm{20}}{{\cal M}_\mathrm{0}}$ &
$\Delta P$         
\\ 
\hline \hline
R100.45.1.4.10 &  1  & 195  & 0.30 & 14.3 & 0.37 & 0.20 & 0.0035 & 750. & 0.032 & 0.12 & 0.32 & 0.24 & 1.02 & 6.9  \rule{0.cm}{0.35cm} \\ 
R100.45.5.4.10 &  5  & 195  & 0.30 & 14.3 & 0.37 & 0.20 & 0.0035 & 150. &  0.16 & 0.38 & 1.02 & 0.74 & 0.89 & 5.3 \\
R100.45.5.8.10 & 3.5 & 220  & 0.34 & 9.9  & 0.23 & 0.20 & 0.0069 & 110. &  0.18 & 0.33 & 1.43 & 0.76 & 0.96 & 7.3 \\
R100.20.1.4.10 &  1  & 116  & 0.38 & 5.5  & 0.24 & 0.04 & 0.0035 & 750. & 0.049 & 0.12 & 0.50 & 0.22 & 1.02 & 5.3 \\
R100.20.5.4.10 &  5  & 116  & 0.38 & 5.5  & 0.24 & 0.04 & 0.0035 & 150. &  0.24 & 0.41 & 1.71 & 0.90 & 0.92 & 4.2 \\
R100.10.5.4.10 &  5  & 82.1 & 0.45 & 2.6  & 0.16 & 0.01 & 0.0035 & 150. &  0.37 & 0.30 & 1.88 & 0.77 & 1.02 & 3.6 \\
\hline
R50.45.10.8.10 & 3.5 & 220  & 0.34 & 9.9  & 0.23 & 0.81 &  0.028 &  53. &  0.18 & 0.32 & 1.39 & 0.52 & 0.93 & 2.6 \\
\hline
R20.20.5.4.10  &  1  & 116  & 0.38 & 5.5  & 0.24 & 1.00 &  0.087 & 150. & 0.045 & 0.11 & 0.46 & 0.14 & 1.02 & 21. \\
R20.20.10.4.10 &  2  & 116  & 0.38 & 5.5  & 0.24 & 1.00 &  0.087 &  75. & 0.098 & 0.18 & 0.75 & 0.20 & 1.00 & 19. \\
R20.20.12.4.10 & 2.5 & 116  & 0.38 & 5.5  & 0.24 & 1.00 &  0.087 &  60. & 0.12  & 0.18 & 0.75 & 0.12 & 1.02 & 20. \\
R20.20.17.4.10 & 3.5 & 116  & 0.38 & 5.5  & 0.24 & 1.00 &  0.087 &  43. & 0.17  & 0.25 & 1.04 & 0.24 & 0.99 & 17. \\
R20.20.25.4.10 &  5  & 116  & 0.38 & 5.5  & 0.24 & 1.00 &  0.087 &  30. & 0.24  & 0.34 & 1.42 & 0.30 & 0.98 & 19. \\
R20.20.25.8.10 & 3.5 & 136  & 0.42 & 3.8  & 0.14 & 1.00 &   0.17 &  21. & 0.30  & 0.18 & 1.39 & 0.88 & 1.02 & 22. \\
R20.20.5.1.10  &  2  & 85.7 & 0.30 & 9.7  & 0.57 & 1.00 &  0.022 & 300. & 0.041 & 0.23 & 0.40 & 0.85 & 0.99 & 13. \\
\hline
R20.10.5.4.10  &  1  & 82.1 & 0.45 & 2.6  & 0.16 & 0.25 &  0.087 & 150. & 0.073 & 0.11 & 0.68 & 0.09 & 1.02 & 19. \\
R20.10.25.4.10 &  5  & 82.1 & 0.45 & 2.6  & 0.16 & 0.25 &  0.087 &  30. & 0.37  & 0.24 & 1.50 & 0.42 & 1.01 & 13. \\
R20.10.2.1.10  &  1  & 56.4 & 0.37 & 4.8  & 0.43 & 0.25 &  0.022 & 600. & 0.027 & 0.20 & 0.47 & 0.19 & 0.99 & 12. \\
R20.10.5.1.10  &  2  & 56.4 & 0.37 & 4.8  & 0.43 & 0.25 &  0.022 & 300. & 0.055 & 0.24 & 0.56 & 0.13 & 0.99 & 11. \\
R20.10.12.1.10 &  5  & 56.4 & 0.37 & 4.8  & 0.43 & 0.25 &  0.022 & 120. & 0.14  & 0.42 & 0.98 & 0.15 & 0.93 & 10. \\
\hline
R20.20.5.4.5   &  1  & 102. & 0.35 & 8.7  & 0.22 & 1.00 &  0.043 & 150. & 0.053 & 0.09 & 0.41 & 0.15 & 1.03 & 19. \\
R20.20.5.4.20  &  1  & 135. & 0.41 & 3.6  & 0.27 & 1.00 &   0.17 & 150. & 0.044 & 0.13 & 0.48 & 0.13 & 1.01 & 25. \\
R20.20.5.4.40  &  1  & 162. & 0.44 & 2.4  & 0.30 & 1.00 &   0.35 & 150. & 0.039 & 0.18 & 0.60 & 0.14 & 1.02 & 32. \\
R20.20.25.4.5  &  5  & 102. & 0.35 & 8.7  & 0.22 & 1.00 &  0.043 &  30. & 0.27  & 0.31 & 1.41 & 0.42 & 0.98 & 17. \\
R20.20.25.4.15 &  5  & 126. & 0.40 & 4.3  & 0.26 & 1.00 &   0.13 &  30. & 0.23  & 0.28 & 1.08 & 0.22 & 0.97 & 17. \\
R20.20.25.4.20 &  5  & 135. & 0.41 & 3.6  & 0.27 & 1.00 &   0.17 &  30. & 0.22  & 0.26 & 0.96 & 0.17 & 0.96 & 20. \\
R20.20.25.4.40 &  5  & 162. & 0.44 & 2.4  & 0.30 & 1.00 &   0.35 &  30. & 0.20  & 0.27 & 0.90 & 0.15 & 0.97 & 24. \\
R20.20.25.8.5  & 3.5 & 117. & 0.38 & 5.9  & 0.13 & 1.00 &  0.087 &  21. & 0.32  & 0.20 & 1.54 & 0.93 & 1.05 & 18. \\
R20.20.25.8.20 & 3.5 & 164. & 0.45 & 2.6  & 0.16 & 1.00 &   0.35 &  21. & 0.26  & 0.17 & 1.06 & 0.27 & 1.02 & 29. \\
R20.20.25.8.40 & 3.5 & 205. & 0.47 & 1.8  & 0.18 & 1.00 &   0.69 &  21. & 0.23  & 0.19 & 1.06 & 0.32 & 1.02 & 39. \\
R20.20.5.1.20  &  2  & 93.9 & 0.32 & 5.7  & 0.61 & 1.00 &  0.043 & 300. & 0.039 & 0.28 & 0.46 & 0.14 & 0.99 & 14. \\
\hline
R20.10.5.4.5   &  1  & 68.2 & 0.42 & 3.8  & 0.14 & 0.25 &  0.043 & 150. & 0.084 & 0.10 & 0.71 & 0.18 & 1.02 & 15. \\
R20.10.5.4.20  &  1  & 102. & 0.47 & 1.8  & 0.18 & 0.25 &   0.17 & 150. & 0.065 & 0.13 & 0.73 & 0.12 & 1.02 & 26. \\
R20.10.25.4.20 &  5  & 102. & 0.47 & 1.8  & 0.18 & 0.25 &   0.17 &  30. & 0.33  & 0.25 & 1.39 & 0.45 & 0.98 & 17. \\
R20.10.25.4.40 &  5  & 133. & 0.49 & 1.4  & 0.22 & 0.25 &   0.35 &  30. & 0.27  & 0.28 & 1.27 & 0.14 & 0.99 & 25. \\
R20.10.12.1.5  &  5  & 49.9 & 0.34 & 7.6  & 0.39 & 0.25 &  0.011 & 120. & 0.15  & 0.46 & 1.18 & 0.29 & 0.94 & 8.8 \\
R20.10.12.1.20 &  5  & 65.5 & 0.40 & 3.0  & 0.47 & 0.25 &  0.043 & 120. & 0.13  & 0.46 & 0.98 & 0.11 & 0.95 & 13. \\
\hline 
\end{tabular}
\end{center}
\label{tab:mod_params}
\end{table*}
\bibliographystyle{apj}
\bibliography{folini.bib}
\end{document}